\documentclass[journal=jpclcd,manuscript=letter,layout=traditional]{achemso}

\usepackage{graphicx}
\usepackage{tabularx}
\usepackage{dcolumn}
\usepackage{longtable}
\usepackage{tensor}
\usepackage{color}
\usepackage{placeins}
\usepackage{bm}
\usepackage{amsmath}
\usepackage{amsfonts}
\usepackage{amssymb}
\usepackage{float}

\graphicspath{{"d:/Group
Vanicek/Desktop/Herzberg-Teller_papers/Herzberg-Teller_TGA/figures/"}{./figures/}{C:/Users/Jiri/Dropbox/Papers/Chemistry_papers/2017/Herzberg-Teller_TGA/NewFigures/}{C:/Users/Jiri/Dropbox/Papers/Chemistry_papers/2017/Herzberg-Teller_TGA/figures/}}
\title{On-the-fly ab initio semiclassical evaluation of absorption spectra of
polyatomic molecules beyond the Condon approximation}
\author{Aur\'elien Patoz}
\author{Tomislav Begu\v{s}i\'{c}}
\author{Ji\v{r}\'i Van\'i\v{c}ek}
\email{jiri.vanicek@epfl.ch.}
\affiliation{Laboratory of Theoretical Physical Chemistry, Institut des Sciences et
Ing\'enierie Chimiques, Ecole Polytechnique F\'ed\'erale de Lausanne (EPFL),
CH-1015, Lausanne, Switzerland}
\date{\today}
\begin{tocentry}
\includegraphics[width=5cm,height=5cm,keepaspectratio]{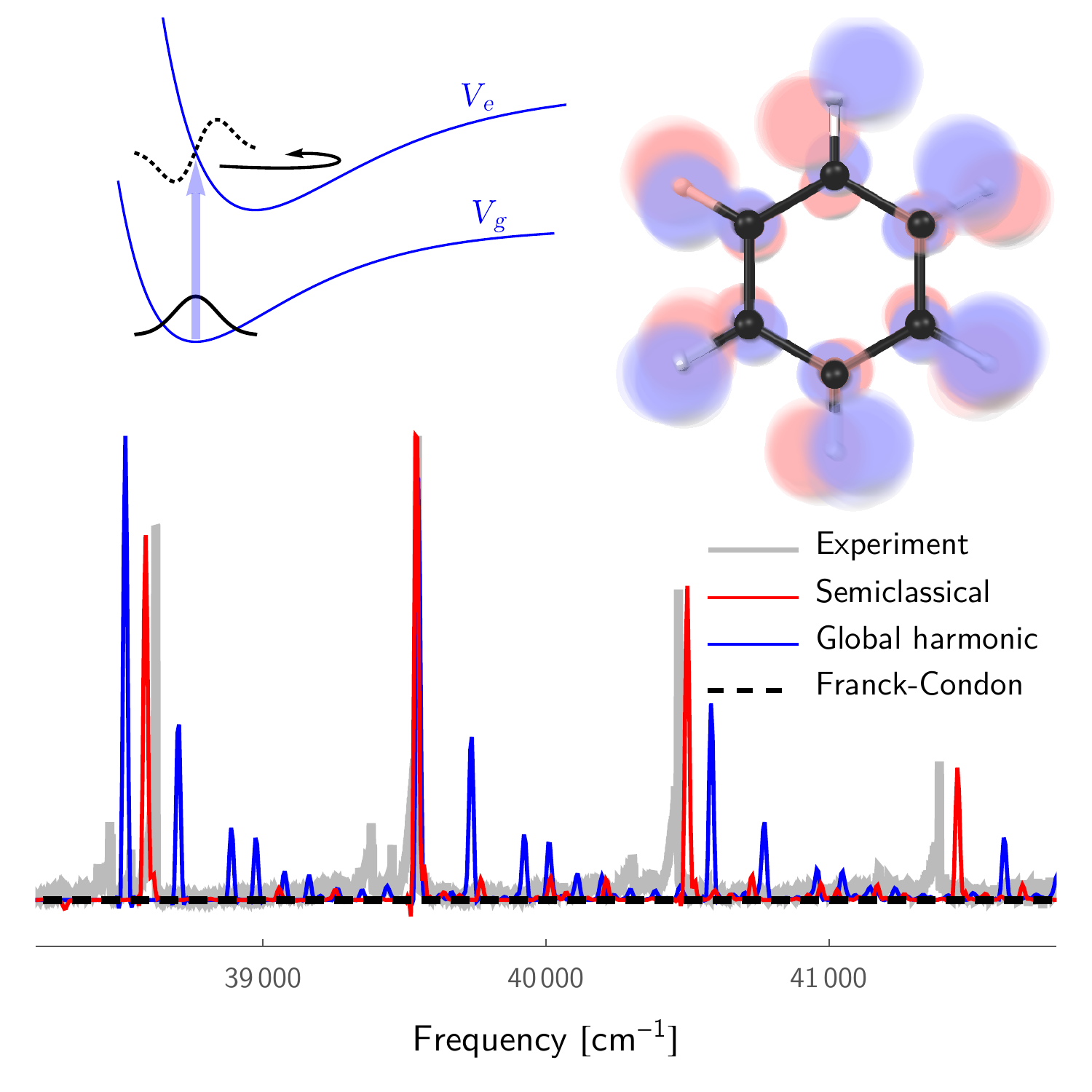}
\end{tocentry}

\begin{document}

\begin{abstract}
To evaluate vibronic spectra beyond the Condon approximation, we extend the
on-the-fly \textit{ab initio} thawed Gaussian approximation by considering the
Herzberg-Teller contribution due to the dependence of the electronic
transition dipole moment on nuclear coordinates. The extended thawed
Gaussian approximation is tested on electronic absorption spectra of phenyl
radical and benzene: Calculated spectra reproduce experimental data and are
much more accurate than standard global harmonic approaches, confirming the
significance of anharmonicity. Moreover, the extended method provides a tool
to quantify the Herzberg-Teller contribution: we show that in phenyl
radical, anharmonicity outweighs the Herzberg-Teller contribution, whereas
in benzene, the Herzberg-Teller contribution is essential, since the
transition is electronically forbidden and Condon approximation yields a
zero spectrum. Surprisingly, both adiabatic harmonic spectra outperform
those of the vertical harmonic model, which describes the Franck-Condon
region better. Finally, we provide a simple recipe for orientationally
averaging spectra, valid beyond Condon approximation, and a relation among
the transition dipole, its gradient, and nonadiabatic coupling vectors.
\end{abstract}

Vibrationally resolved electronic spectroscopy provides a valuable insight into the structure and dynamics of
polyatomic molecules.\cite{Herzberg:1944} Indeed, light-induced molecular
dynamics is recognized as one of the key areas of research in physical
chemistry, not only for fundamental understanding of Nature, but also for
various applications, from solar cells to photodynamic therapy.\cite%
{Albini_Fasani:2017} The development of theoretical methods for simulating
and understanding optical spectra is, therefore, of great importance.

The time-dependent approach to spectroscopy\cite{Heller:1981a} evaluates the
vibronic spectrum as the Fourier transform of the nuclear wavepacket
autocorrelation function, and, in contrast to the commonly used
time-independent Franck-Condon approach,\cite{Condon:1926,Franck_Dymond:1926}
can easily account for effects beyond the Born-Oppenheimer and global
harmonic approximations. In the time-dependent approach, one must first
perform exact or approximate molecular quantum dynamics. While the exact
quantum dynamics typically requires a global potential energy surface\cite%
{Meyer_Worth:2009,Gatti:2014} and thus scales exponentially with dimensions,
semiclassical methods, such as the initial value representation,\cite%
{Miller:1970} thawed Gaussian approximation (TGA),\cite{Heller:1975} frozen
Gaussian approximation,\cite{Heller:1981b} and Herman-Kluk
propagator\cite{Herman_Kluk:1984,Kay:2005} require only local information
and are suitable for on-the-fly implementation. The idea of using multiple
frozen Gaussians as a basis for describing the full wavepacket has inspired
a number of quantum\cite%
{Ben-Nun_Martinez:2000,Saita_Shalashilin:2012,Richings_Lasorne:2015} and
semiclassical \cite%
{Tatchen_Pollak:2009,Ceotto_Atahan:2009a,Ceotto_Atahan:2009b,Wong_Roy:2011,Ianconescu_Pollak:2013,Gabas_Ceotto:2017}
\textquotedblleft first-principles\textquotedblright\ approaches, which
allow a combination with an on-the-fly \textit{ab initio} (OTF-AI)
evaluation of the electronic structure. Apart from a few examples,\cite%
{Grossmann:2006} thawed Gaussians have been largely marginalized since they
can neither describe wavepacket splitting nor very anharmonic dynamics. Yet, 
vibrationally resolved electronic spectra are mostly determined by
short-time dynamics, during which both wavepacket splitting and anharmonic
effects are less important. Indeed, a recent implementation of an on-the-fly 
\textit{ab initio} thawed Gaussian approximation (OTF-AI-TGA) reproduced
successfully vibrational structure of electronic absorption, emission, and
photoelectron spectra, even in rather anharmonic and floppy systems such as
ammonia. \cite{Wehrle_Vanicek:2014,Wehrle_Vanicek:2015}

Here, an extension of the OTF-AI-TGA beyond the Condon approximation is
presented by employing the Herzberg-Teller (HT) approximation,\cite%
{Herzberg_Teller:1933} which, in contrast to the Condon approximation, \cite%
{Condon:1926} includes a linear dependence of the transition dipole moment
on nuclear coordinates. We employ the OTF-AI implementation of the extended
thawed Gaussian approximation (ETGA) \cite{Lee_Heller:1982} to evaluate the
absorption spectra of phenyl radical, an anharmonic system allegedly
exhibiting a significant Herzberg-Teller contribution, \cite%
{Biczysko_Barone:2009} and benzene, a textbook example of a
symmetry-forbidden (i.e., electronically forbidden) transition.\cite%
{Herzberg:1966}

At zero temperature, within the electric-dipole approximation, first-order
time-dependent perturbation theory, and rotating-wave approximation, the
absorption cross section of a molecule with two electronic states that are
not nonadiabatically coupled can be expressed as the Fourier transform 
\begin{equation}
\sigma _{\mu \mu }(\omega )=\frac{2\pi \omega }{\hbar c}\int_{-\infty
}^{\infty }C_{\mu \mu }(t)e^{i\omega t}dt  \label{eq:spectrum}
\end{equation}%
of the dipole time autocorrelation function 
\begin{equation}
C_{\mu \mu }(t)=\langle 1,g|e^{i\hat{H}_{1}t/\hbar }\hat{\mu}e^{-i\hat{H}%
_{2}t/\hbar }\hat{\mu}|1,g\rangle =\langle \phi (0)|\phi (t)\rangle
\,e^{-iE_{1,g}t/\hbar }\,,  \label{eq:correlation}
\end{equation}%
where $\hat{H}_{1}$ and $\hat{H}_{2}$ are nuclear Hamiltonian operators of
the ground and excited electronic states, $\hat{\mu}$ is the matrix element $%
\hat{\vec{\mu}}$ ($=\hat{\vec{\mu}}_{21}$)\footnote{%
The subscript 21 is removed for simplicity since we will almost exclusively
consider this matrix element.} of the electric transition dipole moment
operator projected along the polarization unit vector $\vec{\epsilon}$,
i.e., $\hat{\mu}:=\hat{\vec{\mu}}\cdot \vec{\epsilon}$, and $|1,g\rangle $
is the ground vibrational state of the ground electronic state with
zero-point energy $E_{1,g}$. Thus, the vibronic spectrum can be evaluated by
propagating the initial wavepacket $|\phi (0)\rangle =\hat{\mu}|1,g\rangle $
on the excited state potential energy surface.

To compare with an experiment in gas phase, where the molecules are
isotropically distributed, one must average the computed spectrum over all
molecular orientations. Yet, due to the isotropy of space, a brute-force \
numerical averaging is avoided; averaging is required only over three
orientations of the molecule.\cite{Hein_Rodriguez:2012} To show this,
consider the spectrum and autocorrelation function as $3\times 3$ tensors $%
\sigma _{\vec{\mu}\vec{\mu}}(\omega )$ and $C_{\vec{\mu}\vec{\mu}}(t)$, the
latter related to $C_{\mu \mu }(t)$ by $C_{\mu \mu }(t)=\vec{\epsilon}%
^{\,\top }\cdot C_{\vec{\mu}\vec{\mu}}(t)\cdot \vec{\epsilon}\,.$ A simple
analytical calculation\cite{Begusic_Vanicek:2018} shows that the average
over all orientations of the polarization vector $\vec{\epsilon}$ is $%
\overline{C_{\mu \mu }(t)}=(1/3)\text{Tr}\left[ C_{\vec{\mu}\vec{\mu}}(t)%
\right] $ and the corresponding absorption cross section is 
\begin{equation}
\overline{\sigma (\omega )}=\frac{1}{3}\text{Tr}[\sigma _{\vec{\mu}\vec{\mu}%
}(\omega )]=\frac{1}{3}\left( \sigma _{\mu _{x}\mu _{x}}+\sigma _{\mu
_{y}\mu _{y}}+\sigma _{\mu _{z}\mu _{z}}\right) .
\label{eq:spectrum_averaged}
\end{equation}%
Thus, the average is easily evaluated, e.g., by averaging over only three
arbitrary orthogonal molecular orientations with respect to the fixed
polarization vector $\vec{\epsilon}$ or by fixing the molecular orientation
and averaging over only three arbitrary orthogonal polarization vectors.

While the result (\ref{eq:spectrum_averaged}) holds for arbitrary coordinate
dependence of the transition dipole $\vec{\mu}(q)$, two approximations are
frequently used. Within the most common Condon approximation, the transition
dipole is considered constant: $\vec{\mu}(q)\approx \vec{\mu}(q_{0})$ and
the general result (\ref{eq:spectrum_averaged}) reduces to the textbook
recipe for the averaged Franck-Condon (FC) spectrum:%
\begin{equation}
\overline{\sigma _{\text{FC}}(\omega )}=\frac{1}{3}\sigma _{\left\vert \vec{%
\mu}\right\vert \left\vert \vec{\mu}\right\vert }(\omega ).
\end{equation}%
(\textquotedblleft Divide by $3$ the spectrum for the molecular dipole
aligned with the field.\textquotedblright ) In the more accurate
Herzberg-Teller approximation, the transition dipole moment becomes a linear
function of nuclear coordinates: 
\begin{equation}
\vec{\mu}(q)\approx \vec{\mu}(q_{0})+\text{grad}_{q}\vec{\mu}|_{q_{0}}^{\top
}\cdot (q-q_{0})
\end{equation}%
By explicitly differentiating the matrix element $\vec{\mu}_{\alpha \beta }$
of the molecular dipole between electronic states $\alpha $ and $\beta $
(for a moment we reintroduce the subscripts), one can show\cite%
{Begusic_Vanicek:2018} that 
\begin{equation}
\frac{\partial }{\partial q_{i}}\vec{\mu}_{\alpha \beta }=\sum_{\gamma
}\left( \vec{\mu}_{\alpha \gamma }F_{i,\gamma \beta }-F_{i,\alpha \gamma }%
\vec{\mu}_{\gamma \beta }\right) +\left( e\sum_{j=1}^{N}Z_{j}\frac{\partial 
\vec{R}_{j}}{\partial q_{i}}\right) \delta _{\alpha \beta },
\label{eq:grad_mu_element}
\end{equation}%
or, in a more compact matrix notation:%
\begin{equation}
\frac{\partial }{\partial q_{i}}\vec{\bm{\mu}}=\left[ \vec{\bm{\mu}},\mathbf{%
F}_{i}\right] +\left( e\sum_{j=1}^{N}Z_{j}\frac{\partial \vec{R}_{j}}{%
\partial q_{i}}\right) \mathbf{1},  \label{eq:grad_mu_matrix}
\end{equation}%
where $N$ is the number of atoms, $Z_{j}$ the atomic number and $\vec{R}_{j}$
the coordinates of the $j$th atom, and $F_{i,\alpha \beta }:=\langle \alpha
|(\partial \beta /\partial q_{i})\rangle $ is the $i$th component of the
nonadiabatic coupling vector between states $\alpha $ and $\beta $. For the 
\emph{transition} dipole moment, $\alpha \neq \beta $ and the term
proportional to $\delta _{\alpha \beta }$ in Eq.~(\ref%
{eq:grad_mu_element}) [to $\mathbf{1}$ in Eq.~(\ref%
{eq:grad_mu_matrix})] vanishes, which shows that the Herzberg-Teller
dependence originates from a combination of nonzero nonadiabatic and dipole
couplings of states $\alpha $ and $\beta $ to an intermediate state $\gamma $%
. Moreover, since one may usually neglect nonadiabatic couplings between the
ground and excited electronic states at the ground state optimized geometry,
only the second term of the commutator survives: 
\begin{equation}
\frac{\partial }{\partial q_{j}}\vec{\mu}_{21}\approx -\sum_{\gamma
}F_{i,2\gamma }\vec{\mu}_{\gamma 1}\,,
\end{equation}%
showing that a nonvanishing gradient of the transition dipole between the
ground state $1$ and excited state $2$ requires an intermediate,
\textquotedblleft bright\textquotedblright\ ($\vec{\mu}_{\gamma 1}\neq 0$)
excited state $\gamma $ that is vibronically coupled ($F_{i,2\gamma }\neq 0$%
) to the excited state $2$ . Such interpretation reveals the deep connection
between the Herzberg-Teller approximation and the concepts of
\textquotedblleft vibronic coupling\textquotedblright\ and \textquotedblleft
intensity-borrowing.\textquotedblright \cite{Quack_Merkt:2011} Finally,
although Eq.~(\ref{eq:grad_mu_matrix}) suggests a way to evaluate $\partial
_{q}\vec{\mu}_{\alpha \beta }$ from $\vec{\bm{\mu}}$ and $\mathbf{F}_{i}$,
it is usually easier to evaluate the gradient by finite difference.

To find the autocorrelation function $C_{\mu \mu }(t)$, one must propagate
the wavepacket. Heller's TGA\cite{Heller:1975,Heller:1981a,Lee_Heller:1982}
relies on the fact that a Gaussian wavepacket evolved in at most quadratic
potential remains a Gaussian. Within this approximation, the anharmonicity
of the potential is taken into account partially by propagating the Gaussian
wavepacket 
\begin{equation}
\psi _{t}(q)=N_{0}\exp {\ \left\{ -(q-q_{t})^{\top }\cdot A_{t}\cdot
(q-q_{t})+\frac{i}{\hbar }\left[ p_{t}^{\top }\cdot (q-q_{t})+\gamma _{t}%
\right] \right\} }  \label{eq:GWP}
\end{equation}%
in the time-dependent effective potential given by the local harmonic
approximation of the full potential 
\begin{equation}
V_{\text{eff}}(q,t)=V|_{q_{t}}+(\text{grad}_{q}V|_{q_{t}})^{\top }\cdot
(q-q_{t})+\frac{1}{2}(q-q_{t})^{\top }\cdot \text{Hess}_{q}V|_{q_{t}}\cdot
(q-q_{t})\,,  \label{eq:effPot}
\end{equation}%
where $V|_{q_{t}}$, $\text{grad}_{q}V|_{q_{t}}$, and $\text{Hess}%
_{q}V|_{q_{t}}$ denote the potential energy, gradient, and Hessian evaluated
at the center of the Gaussian, $N_{0}$ is the initial normalization
constant, $(q_{t},p_{t})$ are the phase-space coordinates of the center of
the Gaussian wavepacket at time $t$, $A_{t}$ is the complex symmetric width
matrix, and $\gamma _{t}$ is a complex number; its real part gives an
overall phase factor and its imaginary part ensures the normalization of the
Gaussian wavepacket at all times. Parameters of the Gaussian follow Heller's
equations of motion\cite{Heller:1975,Heller:1981a,Lee_Heller:1982} 
\begin{align}
\dot{q}_{t}& =m^{-1}\cdot p_{t}  \label{eq:dq_dt} \\
\dot{p}_{t}& =-\text{grad}_{q}\,V|_{q_{t}} \\
\dot{A}_{t}& =-2i\hbar A_{t}\cdot m^{-1}\cdot A_{t}+\frac{i}{2\hbar }\text{%
Hess}_{q}V|_{q_{t}} \\
\dot{\gamma}_{t}& =L_{t}-\hbar ^{2}\text{Tr}\left( m^{-1}\cdot A_{t}\right)
\,,  \label{eq:TGA_diff_expressions}
\end{align}%
where $m$ is the diagonal mass matrix and $L_{t}$ the Lagrangian.

The extended TGA used in this work considers a more general form of the
initial wavepacket, namely a Gaussian wavepacket (\ref{eq:GWP}) multiplied
by a polynomial $P\left( q-q_{0}\right) $ in nuclear coordinates, which, at
time zero, can be written as a polynomial in the derivatives with respect to 
$p_{0}$:%
\begin{equation}
\phi _{0}(q)=P\left( q-q_{0}\right) \psi _{0}(q)=P\left( \frac{\hbar }{i}%
\frac{\partial }{\partial p_{0}}\right) \psi _{0}(q)\,.
\end{equation}%
This observation leads to a simple recipe for propagating the extended TGA
wavepacket within the local harmonic approximation;\cite{Lee_Heller:1982}
namely, the wavepacket retains this form at all times: 
\begin{equation}
\phi _{t}(q)=P\left( \frac{\hbar }{i}\frac{\partial }{\partial p_{0}}\right)
\psi _{t}(q),
\end{equation}%
where $\psi _{t}(q)$ is the original TGA wavepacket (\ref{eq:GWP}). As for
the initial Herzberg-Teller wavepacket, $P\left( q-q_{0}\right) =\mu
(q_{0})+b_{0}^{\top }\cdot \left( q-q_{0}\right) $, where $b_{0}=$grad$%
_{q}\mu |_{q_{0}}$, and the semiclassical propagation yields 
\begin{equation}
\phi _{t}(q)=\left[ \mu (q_{0})+b_{t}^{\top }\cdot (q-q_{t})\right] \,\psi
_{t}(q).
\end{equation}%
The four parameters of the Gaussian $\psi _{t}(q)$ are propagated with the
usual TGA equations (\ref{eq:dq_dt})-(\ref{eq:TGA_diff_expressions}) and
only one additional parameter, $b_{t}$, must be evaluated: 
\begin{equation}
b_{t}=\left( -2i\hbar A_{t}\cdot M_{qp,t}+M_{pp,t}\right) \cdot b_{0}\,,
\end{equation}%
where $M_{qp,t}=\partial q_{t}/\partial p_{0}$ and $M_{pp,t}=\partial
p_{t}/\partial p_{0}$ are the elements of the stability matrix, which is
already needed for propagating $A_{t}$ and $\gamma _{t}$, implying that the
additional evaluation of $b_{t}$ comes at almost no additional cost.
Remarkably, the orientational averaging of ETGA spectra is even simpler than
what could be expected from the general simplification we mentioned in the
beginning: within ETGA, the averaging requires only a single trajectory
(instead of three) because the transition dipole moment does not affect the
propagation of $q_{t}$, $p_{t}$, $A_{t}$, $\gamma _{t}$.

As a consequence of this property, the ETGA can be combined with an
on-the-fly \textit{ab initio} scheme at the same cost as the original TGA
for spectra within Condon approximation.\cite%
{Wehrle_Vanicek:2014,Wehrle_Vanicek:2015} The \textit{ab initio}
calculations are typically performed in Cartesian coordinates, and therefore
the \textit{ab initio} gradients and Hessians needed in Eq.~(\ref{eq:effPot}%
) must be transformed\cite{Wehrle_Vanicek:2014,Wehrle_Vanicek:2015} to the
coordinate system $q$ that fits into our framework---the vibrational normal
modes.

As the extended TGA is exact in a globally harmonic potential, it is useful
to compare the on-the-fly approach with two common approximations of the
excited-state potential energy surface: the vertical harmonic (VH) and
adiabatic harmonic (AH)\ approximations,\cite{Domcke_VonNiessen:1977} in
which the excited-state potential is expanded to the second order about the
ground and excited state optimized geometries, respectively (see Refs.~%
\citenum{Wehrle_Vanicek:2014} \ and \citenum{Wehrle_Vanicek:2015} for
details). We use density functional theory for the ground and
time-dependent density functional theory for the excited state \textit{ab initio}
calculations; B3LYP/SNSD for phenyl radical and B3LYP/6-31+G(d,p) for
benzene (see Supporting Information for details and validation by comparison
with a higher level \textit{ab initio} method).

According to Barone and coworkers,\cite%
{Biczysko_Barone:2009,Baiardi_Barone:2013} the calculation of the absorption
spectrum corresponding to $\tilde{\text{A}}^{2}\text{B}_{1}\leftarrow \tilde{%
\text{X}}^{2}\text{A}_{1}$ electronic transition of phenyl radical depends
on the dimensionality of the simulation model and on the inclusion of the
Herzberg-Teller contribution, anharmonicity effects, and mode-mixing
(Duschinsky effect).\cite{Duschinsky:1937} Our model includes all these
effects and provides the means to evaluate their importance.

\begin{figure}[tph]
\includegraphics[scale=1]{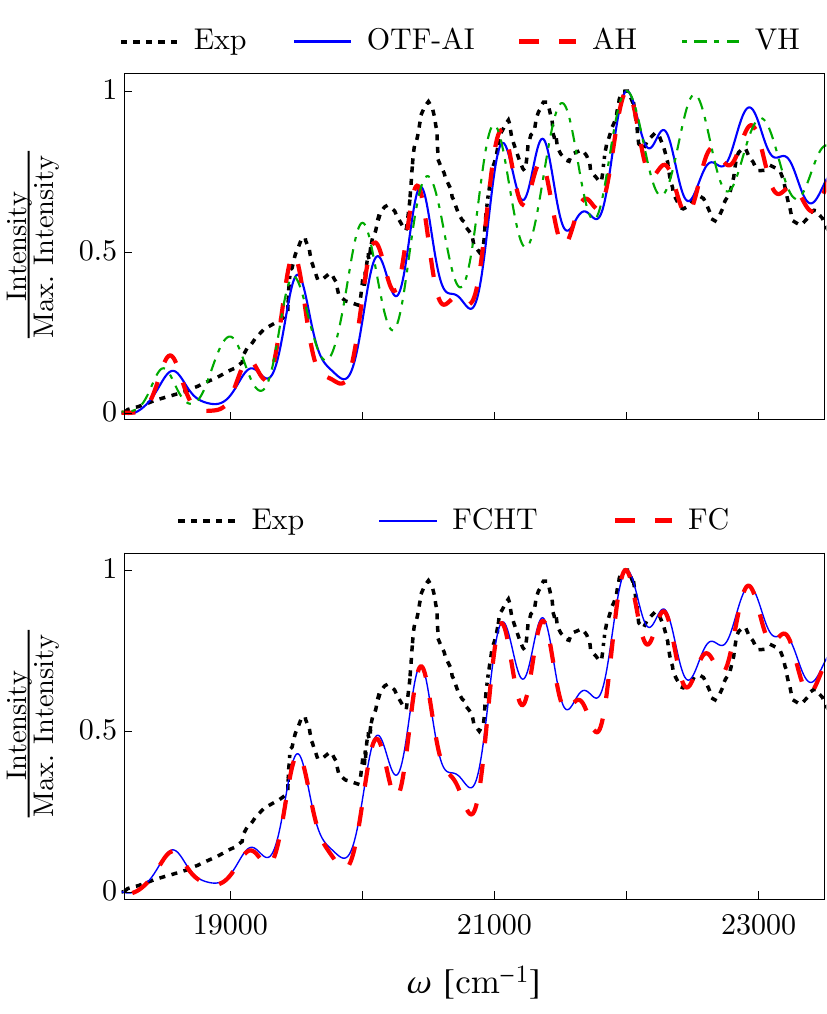}
\caption{Calculated absorption spectra of phenyl radical $\tilde{\text{A}}%
^{2}$B$_{1}\leftarrow \tilde{\text{X}}^{2}$A$_{1}$ electronic transition
compared to the experimental \protect\cite%
{Radziszewski:1999,Baiardi_Barone:2013} spectrum measured in Ar matrix at 6
K. Top: Comparison of the on-the-fly \textit{ab initio} extended TGA
(OTF-AI-ETGA), adiabatic harmonic (AH), and vertical harmonic (VH) models
(all using the FCHT approximation). Bottom: Comparison of the Franck-Condon
(FC) and Franck-Condon Herzberg-Teller (FCHT) approximations (both evaluated
with OTF-AI-ETGA). All spectra are horizontally shifted and rescaled
according to the highest peak (see Table S7 of Supporting Information).}
\label{fig:PhenylRadical_FCHT}
\end{figure}

To assess the influence of anharmonicity, the experimental spectrum is
compared with the spectra simulated using the global harmonic approaches
(Fig.~\ref{fig:PhenylRadical_FCHT}, top). While the vertical harmonic
approach only captures the overall envelope of the experimental spectrum,
but fails to capture any details, the adiabatic harmonic model reproduces
all main features of the spectrum. This is in contrast with the common
expectation that the vertical harmonic approach should be more accurate,\cite%
{Domcke_VonNiessen:1977,Cerezo_Santoro:2013} as it describes better the
Franck-Condon region of the excited state potential. Indeed, the
emission spectra of oligothiophenes\cite{Wehrle_Vanicek:2014} and both
absorption and photoelectron spectra of ammonia\cite{Wehrle_Vanicek:2015}
are much better described with the VH than the AH approach. In phenyl
radical, the failure of the VH approach lies in the incorrect frequencies
and displacements of the two most displaced modes $18$ and $24$ (Fig.~S4 of
Supporting Information), resulting in the missing mode effect:\cite%
{Tannor:2007,Tutt_Heller:1987} when the spectrum is not well resolved, it
may contain a single progression whose spacing does not correspond to any of
the vibrational frequencies of the system. Unlike the global harmonic
approaches, the on-the-fly method overcomes the problem of guessing which
excited state Hessian should be used (i.e., vertical or adiabatic) and
reproduces the experimental spectrum rather well; moreover, with minimum
human input.

To assess the validity of the Condon approximation, the FC and Franck-Condon
Herzberg-Teller (FCHT) spectra of phenyl radical are compared at the bottom
of Fig.~\ref{fig:PhenylRadical_FCHT}. The FC and FCHT\ spectra are very
similar since the absorption spectrum of phenyl radical is mostly determined
by the symmetry-allowed Franck-Condon transition, while the Herzberg-Teller
contribution only broadens the peaks slightly. We find that, in phenyl
radical, including anharmonicity effects with the OTF-AI scheme is more
important than including the HT contribution with the extension of the TGA.

Both the ground and excited state geometries of benzene belong to the $D_{6h}
$ point group. Group theory predicts that the $\tilde{\text{A}}^{1}\text{B}%
_{2\text{u}}\leftarrow \tilde{\text{X}}^{1}\text{A}_{1\text{g}}$ electronic
transition is symmetry-forbidden. Yet it is vibronically allowed, since the
nonzero elements of the gradient of the transition dipole moment give rise
to the vibronic spectrum.\cite{Herzberg:1966} The gradient of the transition
dipole moment originates mostly from nonadiabatic couplings between the $%
\text{B}_{2\text{u}}$ state and the bright $\text{E}_{1\text{u}}$ state.\cite%
{Li_Lin:2010} The spectrum contains one strong progression, assigned to one
of the totally symmetric modes, as well as a number of hot bands. At present
we do not attempt to simulate the hot bands, which require a finite-temperature
treatment;\cite{He_Pollak:2001} our goal is computing
the correct absorption cross sections of the main progression. In addition,
we do not treat the spectral features arising from the second order vibronic
coupling, i.e., a number of small intensity peaks which cannot be described
within the first-order Herzberg-Teller approximation.\cite{Fischer_Ross:1981}

\begin{figure*}[tph]
\includegraphics[scale = 1]{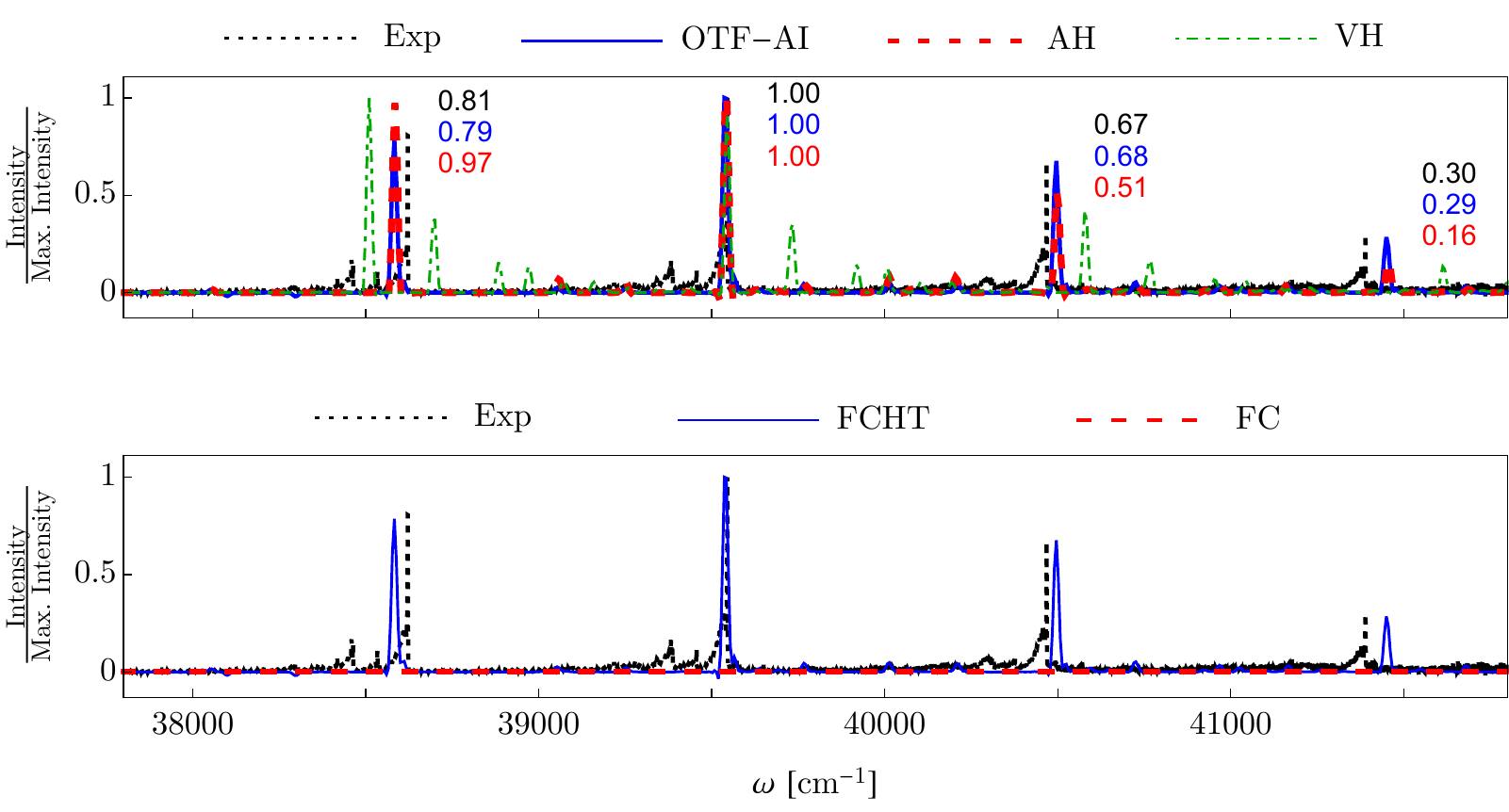}
\caption{Calculated absorption spectra of benzene $\tilde{\text{A}}^{1}$B$_{2%
\text{u}}\leftarrow \tilde{\text{X}}^{1}$A$_{1\text{g}}$ electronic
transition compared to the experimental \protect\cite%
{Fally_Vandaele:2009,Keller-Rudek_Sorensen:2013} spectrum measured at 293 K.
See the caption of Fig.~\protect\ref{fig:PhenylRadical_FCHT} for details. 
To clarify the difference between the adiabatic harmonic (AH, red)
and on-the-fly (blue) spectra, we show the scaled intensities of the
experimental, AH, and on-the-fly peaks.}
\label{fig:Benzene_FCHT__FC_Exp}
\end{figure*}

\begin{figure*}[tph]
\centering \includegraphics[scale=1]{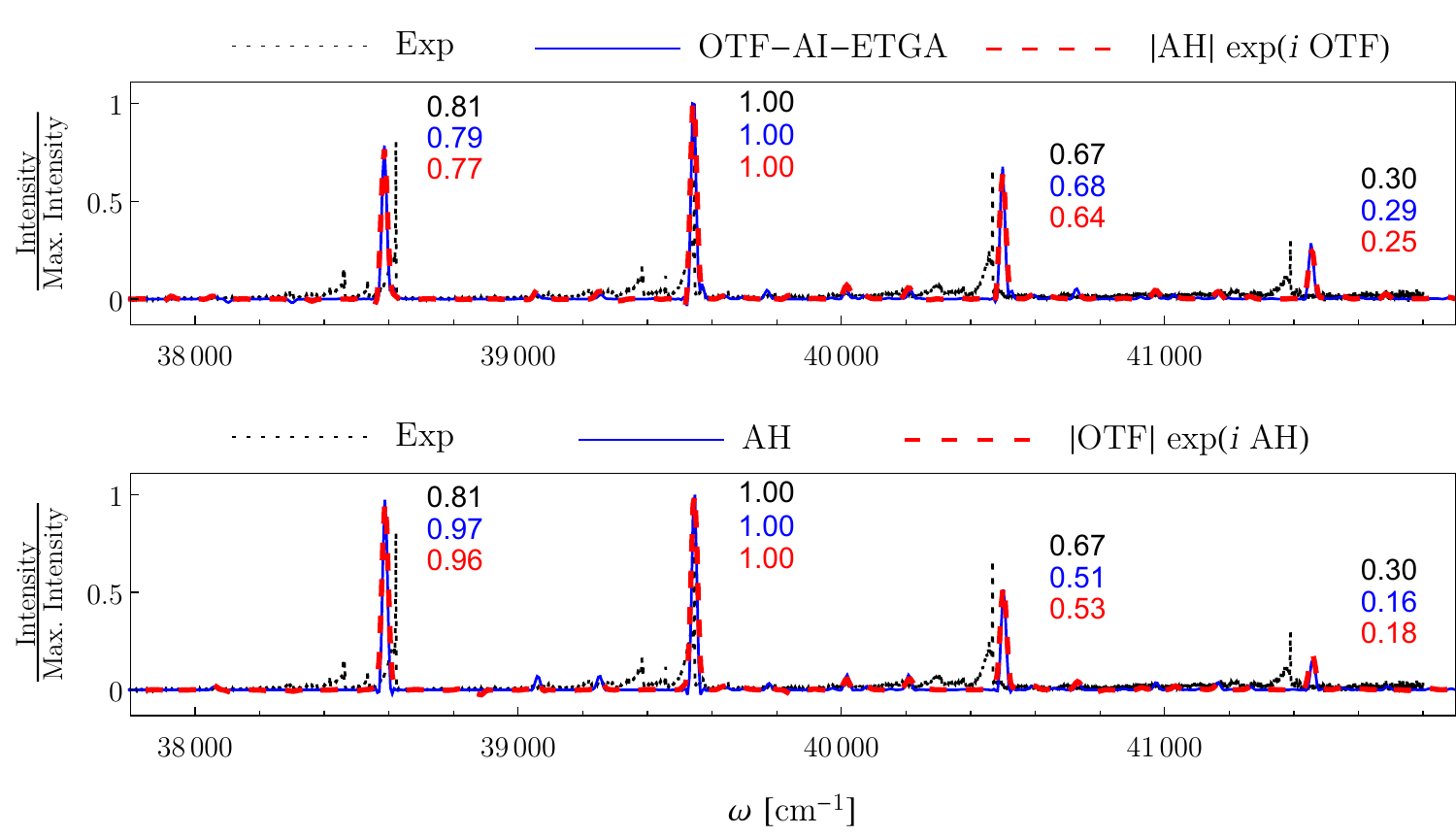}
\caption{Spectra calculated from the hybrid autocorrelation functions: the $|%
\text{AH}|~\exp (i\,\text{OTF})$ spectrum is almost the same as the very
accurate OTF-AI-ETGA spectrum (top), while the $|\text{OTF}|~\exp (i\,\text{%
AH})$ spectrum overlaps with the less accurate adiabatic harmonic spectrum
(bottom), confirming that the error in the phase of the autocorrelation
function is more important than the error in its magnitude (see Table S7 of
Supporting Information). Scaled peak intensities are shown as in
Fig.~\protect\ref{fig:Benzene_FCHT__FC_Exp}.}
\label{fig:hybrid}
\end{figure*}

Interestingly, the adiabatic harmonic approach again reproduces the
experimental spectrum at least qualitatively, unlike the vertical
harmonic approach which results in a number of peaks not observed in the
experiment (Fig.~\ref{fig:Benzene_FCHT__FC_Exp}, top). In the VH model,
modes 25, 29 and 30 are more distorted since their frequencies are
significantly lower than the corresponding frequencies in the AH model (see
Table~S6 and Fig.~S5 of Supporting Information). Again, due to a partial
treatment of anharmonicity, the OTF-AI-ETGA spectrum shows significant
improvement over both global harmonic methods: While the relative
intensities of AH peaks have errors of $20$ to $50\%$ and the VH model fails completely, the relative intensities of the
OTF-AI-ETGA peaks lie within $5\%$ of experiment. Influence of anharmonicity on the spectrum is
investigated using the autocorrelation functions in Supporting Information
(see Figs. S6--S8). To further explore whether it is the error in the phase
or magnitude of the autocorrelation function that affects the spectra more,
we construct two hybrid, nonphysical autocorrelation functions. The first,
denoted $|\text{AH}|~\exp (i\,\text{OTF})$, combines the magnitude of the
adiabatic harmonic autocorrelation function with the phase from the
on-the-fly correlation function, while the second, denoted $|\text{OTF}%
|~\exp (i\,\text{AH})$, combines the magnitude of the on-the-fly
autocorrelation function with the phase from the adiabatic harmonic
autocorrelation function. The spectra in Fig.~\ref{fig:hybrid} simulated
using these two hybrid autocorrelations show clearly that it is the error in
the phase of the AH model that most corrupts the intensities of the peaks.
The difference between the two phases is closely related to the propagation
of the stability matrix, and, consequently, to the Hessians of the excited
electronic state potential.

In contrast to almost perfect description of the intensities by the
OTF-AI-ETGA, neither the AH nor the OTF-AI-ETGA reproduces correctly the
experimental spacing of the peaks in the main progression. This error,
however, can be assigned to the electronic structure method, implying that
the use of other density functional or wavefunction-based methods could give
more accurate curvature of the potential, and hence a better spacing.

Our main result is contained in the bottom panel of Fig.~\ref%
{fig:Benzene_FCHT__FC_Exp}, which compares the Franck-Condon spectrum (based
on the Condon approximation), which can be evaluated with the original TGA,
and the FCHT spectrum (based on the Herzberg-Teller approximation), which
requires the extended TGA. While the FCHT agrees very well with experiment,
the Franck-Condon spectrum is zero, since the transition dipole moment at
the ground-state equilibrium is zero, which is the precise meaning of an
electronically forbidden transition.

Incidentally, one can imagine simulating the FC spectrum by the commonly
adopted procedure in which the transition dipole moment is set to unity and
the spectrum is rescaled at the end. Although reasonable for electronically
allowed transitions, here this \textquotedblleft blind\textquotedblright\
procedure makes no sense since $\mu _{\text{FC}}=0$. This \textquotedblleft
FC\textquotedblright\ approach is compared to the FCHT result (see Fig.~S9
of Supporting Information), and, somewhat surprisingly, the spectra are very
similar, the main difference being a constant shift corresponding exactly to
the excited state frequency of the degenerate inducing modes 27 and 28 (see
Tables S5--S6 of Supporting Information). However, such similarity between
the \textquotedblleft FC\textquotedblright\ and the FCHT approach is not
general. In addition, the \textquotedblleft FC\textquotedblright\ approach
is not capable of reproducing the \emph{absolute} magnitudes of absorption
cross sections, while the full FCHT approach provides a good estimate of the
absolute absorption cross sections, as shown in Fig.~S10 of Supporting
Information.

To conclude, we presented an extension to the on-the-fly \textit{ab initio}
TGA and employed it to evaluate absorption spectra of phenyl radical and
benzene within the Herzberg-Teller approximation. Considerable improvement
compared to the usual global harmonic approaches was achieved and a further
insight into the origins of the spectral features was given. The results
obtained for the absorption spectrum of phenyl radical imply that including
the anharmonicity effects is more important than the Herzberg-Teller
contribution to the spectrum. Although thawed Gaussian approximation is
often described in the context of calculating low-resolution spectra, here
we reported the evaluation of a rather high-resolution absorption spectrum
of benzene with high accuracy. The improvement was especially pronounced in
the intensities of the peaks, due to a partial inclusion of the
anharmonicity of the excited state potential. Thus, the OTF-AI-ETGA can be
used not only to reproduce spectra, but more importantly, to evaluate the
importance of different effects by going beyond the commonly used Condon and
global harmonic approximations.

\begin{suppinfo}
Computational details, ground and excited state optimized geometries, transition dipole moments, frequencies, 
and displacements, validation of the Born-Oppenheimer 
approximation and of the choice of the electronic structure method, 
conservation of the phase space volume and symplectic structure,
details of spectra calculations, models describing the failure of the vertical harmonic method for absorption spectra, 
discussion of anharmonicity effects in benzene, ``Franck-Condon'' (``FC'') spectrum, and absolute absorption 
cross sections of benzene.
\end{suppinfo}

\begin{acknowledgement}
The authors acknowledge the financial support from the Swiss National Science Foundation through the NCCR 
MUST (Molecular Ultrafast Science and Technology) Network, from the European Research Council (ERC) under 
the European Union's Horizon 2020 research and innovation programme (grant agreement No. 683069 -- 
MOLEQULE), and from the EPFL.
\end{acknowledgement}

\bibliography{Herzberg-Teller_TGA}

\end{document}


\graphicspath{{"d:/Group Vanicek/Desktop/Herzberg-Teller_papers/Herzberg-Teller_TGA/figures/"}{./figures/}
{C:/Users/Jiri/Dropbox/Papers/Chemistry_papers/2017/Herzberg-Teller_TGA/NewFigures/}
{C:/Users/Jiri/Dropbox/Papers/Chemistry_papers/2017/Herzberg-Teller_TGA/figures/}}

\newpage

\section{\label{sec:compdet}Computational details}

All ab initio calculations were performed using GAUSSIAN09\cite{g09}
electronic structure package. Density functional theory and time-dependent
density functional theory have been used with the B3LYP functional for
ground and excited states, respectively. 6-31+G(d,p) basis set was used for
benzene, while for the phenyl radical we used the SNSD basis set---same basis
set was used with B3LYP functional in a previous study.\cite%
{Baiardi_Barone:2013} An in-house code for running OTF-AI classical dynamics
was used in combination with GAUSSIAN09 to compute the trajectories for both
the phenyl radical and benzene with a time step of 8 a.u. ($\approx $ 0.2 fs).
Hessians were then computed at every fourth step of the trajectory (and
interpolated in between).\cite{Wehrle_Vanicek:2014} The OTF-AI trajectories
were run for $T\approx 580$ fs (3000 steps) in the case of the phenyl radical
and $T\approx 2$ ps ($10^{4}$ steps) for benzene. Due to numerical
computation of the excited-state Hessians, the computational cost should be
roughly $6N$ times larger than the cost of the trajectory
calculation (each Hessian calculation consists of $6N$ gradient
evaluations), where $N$ is the number of atoms. However, this is
relaxed in two ways: the Hessians are not computed at each step of the
trajectory and they can be computed in parallel (Hessian information is not
needed for the trajectory propagation). As a result, for both the phenyl radical and benzene, the Hessian
calculations took approximately as much real time as the \textit{ab initio} trajectory
calculation.

The wavepacket propagation was performed in the ground-state
mass-scaled normal mode coordinates; the Hessians, forces, and the gradient
of the transition dipole moment were transformed from Cartesian to
mass-scaled normal mode coordinates by following the prescription explained
in detail in Refs.~\citenum{Wehrle_Vanicek:2014} and %
\citenum{Wehrle_Vanicek:2015}. The initial Gaussian wavepacket was chosen to
be the ground vibrational state of the harmonic fit to the ground potential
energy surface. Since only a single set of coordinates is used to represent
all quantities, there is no need for additional corrections to account for
mode-mixing (Duschinsky effect)---mode-mixing is reflected in the
non-diagonal elements of the excited-state Hessian matrix represented in the
ground-state normal modes.

The gradient of the transition dipole moment was computed using the
first-order finite differences scheme with a step of $10^{-2}$ \AA ~for
the phenyl radical and $10^{-4}$ \AA ~for benzene. For the phenyl radical, larger
step was used; at smaller steps the gradient depends significantly on the
step size, whereas at larger step sizes, these values vary only very slowly.
The dependence is likely to arise from inaccuracy of the transition dipole
moments.

Orientational averaging was performed by computing the autocorrelation
functions for three orthogonal orientations of the electric field (i.e. by
considering three orthogonal elements of the transition dipole moment $\mu_x$%
, $\mu_y$, and $\mu_z$) and then taking their average. Since it is only the
polynomial part of the wavepacket that changes with the transition dipole
moment, the Gaussian parameters of the wavepacket---its position, momentum,
width, and phase---are not affected when switching the polarization of the
electric field; the on-the-fly \textit{ab initio} extended TGA still
requires only one classical trajectory. For the phenyl radical, all three
elements of the transition dipole moment are non-zero, whereas for benzene,
it is only $x$ and $y$ polarizations which give rise to the observed
spectra---both the transition dipole moment and its gradient are zero for $z$
polarization. Thus, for benzene, only two correlation functions have to be
evaluated for orientational averaging.

The autocorrelation functions were evaluated for positive times and the
spectra were computed numerically using the fast Fourier transform. The
broadening of the spectra was introduced by damping the autocorrelation
functions: for the phenyl radical a Gaussian decay corresponding to a Gaussian
broadening of the spectrum with half-width at half-maximum of 100 cm$^{-1}$
was used, while for the well-resolved benzene spectrum, $\Gamma (t)=\text{cos%
}^{2}(\frac{\pi t}{2T})$ damping function was used.

\section{Ground- and excited-state optimized geometries, transition dipole
moments, frequencies, and displacements}

\subsection{Phenyl radical}

\begin{table}[H]
\caption{Optimized geometries (in \AA ) of ground ($\tilde{\text{X}}^2\text{A%
}_1$) and first excited ($\tilde{\text{A}}^2\text{B}_1$) electronic states
of the phenyl radical.}
\label{tab:PhenylRadical_optgeom}\centering
\begin{tabular}{ccccccccc}
\hline\hline
\multicolumn{3}{r}{$\tilde{\text{X}}^2\text{A}_1$} &  &  & 
\multicolumn{3}{r}{$\tilde{\text{A}}^2\text{B}_1$} &  \\ 
& X & Y & Z &  &  & X & Y & Z \\ \hline
C & 0.000 & 1.227 & 0.772 &  &  & 0.000 & 1.213 & 0.734 \\ 
C & 0.000 & 1.214 & -0.632 &  &  & 0.000 & 1.228 & -0.644 \\ 
C & 0.000 & 0.000 & -1.324 &  &  & 0.000 & 0.000 & -1.343 \\ 
C & 0.000 & -1.214 & -0.632 &  &  & 0.000 & -1.228 & -0.644 \\ 
C & 0.000 & -1.227 & 0.772 &  &  & 0.000 & -1.213 & 0.734 \\ 
C & 0.000 & 0.000 & 1.398 &  &  & 0.000 & 0.000 & 1.549 \\ 
H & 0.000 & 2.163 & 1.324 &  &  & 0.000 & 2.169 & 1.258 \\ 
H & 0.000 & -2.154 & -1.178 &  &  & 0.000 & -2.160 & -1.204 \\ 
H & 0.000 & -2.163 & 1.324 &  &  & 0.000 & -2.169 & 1.258 \\ 
H & 0.000 & 2.154 & -1.178 &  &  & 0.000 & 2.160 & -1.204 \\ 
H & 0.000 & 0.000 & -2.410 &  &  & 0.000 & 0.000 & -2.430 \\ \hline\hline
\end{tabular}%
\end{table}

\begin{table}[H]
\caption{Derivatives of the transition dipole moment (in atomic units) of
the phenyl radical $\tilde{\text{A}}^2\text{B}_1 \leftarrow \tilde{\text{X}}%
^2\text{A}_1$ electronic transition with respect to the mass-scaled normal
mode coordinates of the ground electronic state. The transition dipole
moment elements evaluated at the ground-state optimized geometry are: $\protect%
\mu_x = 0.1354\, \text{a.u.} = 0.344\, \text{D}$, $\protect\mu_y=\protect\mu%
_z=0$.}
\label{tab:PhenylRadical_TDM}\centering
\begin{tabular}{cccccccccc}
\hline\hline
Mode &  & Symmetry &  & $\partial \mu_x / \partial q$ &  & $\partial \mu_y /
\partial q$ &  & $\partial \mu_z / \partial q$ &  \\ \hline
1 &  & a$_2$ &  & 0 &  & 0.00131 &  & 0 &  \\ 
2 &  & b$_1$ &  & 0 &  & 0 &  & 0.00207 &  \\ 
3 &  & b$_2$ &  & 0 &  & 0 &  & 0 &  \\ 
4 &  & a$_1$ &  & -0.00021 &  & 0 &  & 0 &  \\ 
5 &  & b$_1$ &  & 0 &  & 0 &  & -0.00404 &  \\ 
6 &  & b$_1$ &  & 0 &  & 0 &  & 0.00231 &  \\ 
7 &  & a$_2$ &  & 0 &  & -0.00026 &  & 0 &  \\ 
8 &  & b$_1$ &  & 0 &  & 0 &  & -0.00068 &  \\ 
9 &  & a$_2$ &  & 0 &  & 0.00124 &  & 0 &  \\ 
10 &  & a$_1$ &  & 0.00057 &  & 0 &  & 0 &  \\ 
11 &  & b$_1$ &  & 0 &  & 0 &  & 0.00201 &  \\ 
12 &  & a$_1$ &  & 0.00001 &  & 0 &  & 0 &  \\ 
13 &  & a$_1$ &  & -0.00017 &  & 0 &  & 0 &  \\ 
14 &  & b$_2$ &  & 0 &  & 0 &  & 0 &  \\ 
15 &  & b$_2$ &  & 0 &  & 0 &  & 0 &  \\ 
16 &  & a$_1$ &  & 0.00000 &  & 0 &  & 0 &  \\ 
17 &  & b$_2$ &  & 0 &  & 0 &  & 0 &  \\ 
18 &  & b$_2$ &  & 0 &  & 0 &  & 0 &  \\ 
19 &  & b$_2$ &  & 0 &  & 0 &  & 0 &  \\ 
20 &  & a$_1$ &  & -0.00071 &  & 0 &  & 0 &  \\ 
21 &  & a$_1$ &  & 0.00050 &  & 0 &  & 0 &  \\ 
22 &  & b$_2$ &  & 0 &  & 0 &  & 0 &  \\ 
23 &  & a$_1$ &  & 0.00024 &  & 0 &  & 0 &  \\ 
24 &  & b$_2$ &  & 0 &  & 0 &  & 0 &  \\ 
25 &  & a$_1$ &  & 0.00041 &  & 0 &  & 0 &  \\ 
26 &  & b$_2$ &  & 0 &  & 0 &  & 0 &  \\ 
27 &  & a$_1$ &  & 0.00028 &  & 0 &  & 0 &  \\ \hline\hline
\end{tabular}%
\end{table}

\begin{table}[H]
\caption{Normal modes of the phenyl radical ground electronic state with
corresponding frequencies (in $\text{cm}^{-1}$). The excited-state
frequencies $\protect\omega$ computed at the optimized ground-state geometry
[vertical harmonic (VH) model] and optimized excited-state geometry
[adiabatic harmonic (AH) model] are given in $\text{cm}^{-1}$. The
displacements $\protect\delta$ between the minima of the ground and excited
state global harmonic potentials using mass-scaled normal mode coordinates
are given in atomic units (1~a.u.~$\simeq 0.0123943 \,\protect\sqrt{\text{amu%
}}\,\text{\AA }$). Relative displacements $\Delta = A_0^{1/2} \cdot \protect%
\delta$, where $A_0$ is the width matrix of the initial wavepacket, are
dimensionless and more appropriate measures of excitation since they take
into account the fact that a displacement $\protect\delta$ of low frequency
mode has a smaller effect on the resulting spectrum than the same
displacement $\protect\delta$ of the high-frequency mode. Normal modes are
ordered from the highest to the lowest ground-state frequency.}
\label{tab:PhenylRadical_freq}\centering
\begin{tabular}{ccccccccccccc}
\hline\hline
& Mode & Symmetry & Ground state &  & \multicolumn{3}{c}{Excited state AH} & %
\phantom{abc} & \multicolumn{3}{c}{Excited state VH} &  \\ 
&  &  & $\omega$ &  & $\omega$ & $\delta$ & $\Delta$ &  & $\omega$ & $\delta$ & $%
\Delta$ &  \\ 
&  &  & cm$^{-1}$ &  & cm$^{-1}$ & a.u. &  &  & cm$^{-1}$ & a.u. &  &  \\ 
\hline
& 1 & a$_1$ & 3193 &  & 3187 & -0.06 & -0.0054 &  & 3195 & 0.01 & 0.0007 & 
\\ 
& 2 & b$_2$ & 3184 &  & 3168 & 0 & 0 &  & 3166 & 0 & 0 &  \\ 
& 3 & a$_1$ & 3181 &  & 3162 & -0.14 & -0.0122 &  & 3167 & -0.04 & -0.0032 & 
\\ 
& 4 & b$_2$ & 3168 &  & 3140 & 0 & 0 &  & 3155 & 0 & 0 &  \\ 
& 5 & a$_1$ & 3161 &  & 3139 & 0.03 & 0.0025 &  & 3150 & -0.02 & -0.0017 & 
\\ 
& 6 & b$_2$ & 1631 &  & 1522 & 0 & 0 &  & 1548 & 0 & 0 &  \\ 
& 7 & a$_1$ & 1573 &  & 1633 & 12.92 & 0.7732 &  & 1571 & 13.96 & 0.8357 & 
\\ 
& 8 & a$_1$ & 1468 &  & 1443 & -9.41 & -0.5442 &  & 1447 & -10.39 & -0.6007
&  \\ 
& 9 & b$_2$ & 1460 &  & 1395 & 0 & 0 &  & 1494 & 0 & 0 &  \\ 
& 10 & b$_2$ & 1336 &  & 1351 & 0 & 0 &  & 1364 & 0 & 0 &  \\ 
& 11 & b$_2$ & 1303 &  & 1243 & 0 & 0 &  & 1273 & 0 & 0 &  \\ 
& 12 & a$_1$ & 1172 &  & 1214 & -3.39 & -0.1749 &  & 1192 & -3.36 & -0.1890
&  \\ 
& 13 & b$_2$ & 1171 &  & 1119 & 0 & 0 &  & 1090 & 0 & 0 &  \\ 
& 14 & b$_2$ & 1069 &  & 1049 & 0 & 0 &  & 1037 & 0 & 0 &  \\ 
& 15 & a$_1$ & 1047 &  & 1018 & -11.51 & -0.5623 &  & 1071 & -9.87 & -0.4820
&  \\ 
& 16 & a$_1$ & 1015 &  & 1000 & 7.79 & 0.3745 &  & 1023 & 4.65 & 0.2237 & 
\\ 
& 17 & b$_1$ & 997 &  & 1027 & 0 & 0 &  & 1014 & 0 & 0 &  \\ 
& 18 & a$_1$ & 981 &  & 924 & 30.81 & 1.4563 &  & 950 & 33.80 & 1.5977 &  \\ 
& 19 & a$_2$ & 970 &  & 998 & 0 & 0 &  & 971 & 0 & 0 &  \\ 
& 20 & b$_1$ & 895 &  & 968 & 0 & 0 &  & 934 & 0 & 0 &  \\ 
& 21 & a$_2$ & 815 &  & 788 & 0 & 0 &  & 800 & 0 & 0 &  \\ 
& 22 & b$_1$ & 721 &  & 768 & 0 & 0 &  & 696 & 0 & 0 &  \\ 
& 23 & a$_1$ & 671 &  & 685 & 0 & 0 &  & 673 & 0 & 0 &  \\ 
& 24 & a$_1$ & 614 &  & 590 & -28.08 & -1.0498 &  & 541 & -39.27 & -1.4682 & 
\\ 
& 25 & b$_2$ & 593 &  & 528 & 0 & 0 &  & 419 & 0 & 0 &  \\ 
& 26 & b$_1$ & 425 &  & 355 & 0 & 0 &  & 223 & 0 & 0 &  \\ 
& 27 & a$_2$ & 401 &  & 301 & 0 & 0 &  & 212 & 0 & 0 &  \\ \hline\hline
\end{tabular}%
\end{table}

\subsection{Benzene}

\begin{table}[H]
\caption{Bond lengths (in \AA ) at optimized geometries for ground ($\tilde{%
\text{X}}^1\text{A}_{1\text{g}}$) and first excited ($\tilde{\text{A}}^1%
\text{B}_{2\text{u}}$) states of benzene.}
\label{tab:Benzene_optgeom}\centering
\begin{tabular}{@{}ccccc}
\hline\hline
&  & $\tilde{\text{X}}^1\text{A}_{1\text{g}}$ &  & $\tilde{\text{A}}^1\text{B%
}_{2\text{u}}$ \\ \hline
$R_{\text{C-C}}$ &  & 1.398 &  & 1.430 \\ 
$R_{\text{C-H}}$ &  & 1.086 &  & 1.085 \\ \hline\hline
\end{tabular}%
\end{table}

\begin{table}[]
\caption{Derivatives of the transition dipole moment (in atomic units) of
the benzene $\tilde{\text{A}}^1\text{B}_{2\text{u}} \leftarrow \tilde{\text{X%
}}^1\text{A}_{1\text{g}}$ electronic transition with respect to the normal
mode coordinates associated with doubly-degenerate normal modes that
transform as $e_{2g}$ irreducible representation. These modes are called the 
\emph{inducing modes}, as they are responsible for the observation of the
formally symmetry-forbidden electronic transition in benzene. The strongest
transitions are induced by the low-frequency modes 27 and 28, which exhibit
the largest derivatives of the transition dipole moment. The molecule is
oriented according to Mulliken's convention (standard orientation): the axis
of greatest symmetry is aligned with the $z$-axis, molecule lies in the $xy$
plane, and the $y$-axis passes through two hydrogen and two carbon atoms.
All elements of the transition dipole moment and its gradient that are not
listed below are zero.}
\label{tab:Benzene_TDM}\centering
\begin{tabular}{ccccccccc}
\hline\hline
Mode &  &  & $\partial \mu_x / \partial q$ &  & $\partial \mu_y / \partial q$
&  & $\partial \mu_z / \partial q$ &  \\ \hline
4 &  &  & 0.00055 &  & -0.00265 &  & 0 &  \\ 
5 &  &  & -0.00207 &  & -0.00071 &  & 0 &  \\ 
7 &  &  & -0.00002 &  & -0.00242 &  & 0 &  \\ 
8 &  &  & 0.00006 &  & -0.00064 &  & 0 &  \\ 
13 &  &  & -0.00036 &  & 0.00067 &  & 0 &  \\ 
14 &  &  & 0.00133 &  & 0.00018 &  & 0 &  \\ 
27 &  &  & 0.00784 &  & 0.00235 &  & 0 &  \\ 
28 &  &  & 0.00208 &  & -0.00883 &  & 0 &  \\ \hline\hline
\end{tabular}%
\end{table}

\begin{table}[H]
\caption{Normal modes of benzene ground state with corresponding frequencies
(in $\text{cm}^{-1}$), excited-state frequencies $\protect\omega$ computed
at the optimized ground-state geometry [vertical harmonic (VH) model] and
optimized excited-state geometry [adiabatic harmonic (AH) model] (in $\text{%
cm}^{-1}$), displacements $\protect\delta$ between the minima of the ground
state and excited-state global harmonic potentials (in atomic units), and
dimensionless relative displacements $\Delta = A_0^{1/2} \cdot \protect%
\delta $. See the caption of Table~\protect\ref{tab:PhenylRadical_freq} for
details. }
\label{tab:Benzene_freq}\centering
\begin{tabular}{ccccccccccccc}
\hline\hline
& Mode & Symmetry & Ground state &  & \multicolumn{3}{c}{Excited state AH} & %
\phantom{abc} & \multicolumn{3}{c}{Excited state VH} &  \\ 
&  &  & $\omega$ &  & $\omega$ & $\delta$ & $\Delta$ &  & $\omega$ & $\delta$ & $%
\Delta$ &  \\ 
&  &  & cm$^{-1}$ &  & cm$^{-1}$ & a.u. &  &  & cm$^{-1}$ & a.u. &  &  \\ 
\hline
& 1 & a$_{1\text{g}}$ & 3207 &  & 3232 & 0.90 & 0.077 &  & 3212 & 0.87 & 
0.074 &  \\ 
& 2 & e$_{1\text{u}}$ & 3197 &  & 3221 & 0 & 0 &  & 3199 & 0 & 0 &  \\ 
& 3 & e$_{1\text{u}}$ & 3197 &  & 3221 & 0 & 0 &  & 3199 & 0 & 0 &  \\ 
& 4 & e$_{2\text{g}}$ & 3182 &  & 3207 & 0 & 0 &  & 3183 & 0 & 0 &  \\ 
& 5 & e$_{2\text{g}}$ & 3182 &  & 3207 & 0 & 0 &  & 3183 & 0 & 0 &  \\ 
& 6 & b$_{1\text{u}}$ & 3172 &  & 3201 & 0 & 0 &  & 3177 & 0 & 0 &  \\ 
& 7 & e$_{2\text{g}}$ & 1642 &  & 1571 & 0 & 0 &  & 1678 & 0 & 0 &  \\ 
& 8 & e$_{2\text{g}}$ & 1642 &  & 1571 & 0 & 0 &  & 1678 & 0 & 0 &  \\ 
& 9 & e$_{1\text{u}}$ & 1515 &  & 1449 & 0 & 0 &  & 1497 & 0 & 0 &  \\ 
& 10 & e$_{1\text{u}}$ & 1515 &  & 1449 & 0 & 0 &  & 1497 & 0 & 0 &  \\ 
& 11 & a$_{2\text{g}}$ & 1378 &  & 1361 & 0 & 0 &  & 1385 & 0 & 0 &  \\ 
& 12 & b$_{2\text{u}}$ & 1353 &  & 1474 & 0 & 0 &  & 1620 & 0 & 0 &  \\ 
& 13 & e$_{2\text{g}}$ & 1198 &  & 1179 & 0 & 0 &  & 1215 & 0 & 0 &  \\ 
& 14 & e$_{2\text{g}}$ & 1198 &  & 1179 & 0 & 0 &  & 1215 & 0 & 0 &  \\ 
& 15 & b$_{2\text{u}}$ & 1176 &  & 1175 & 0 & 0 &  & 1201 & 0 & 0 &  \\ 
& 16 & e$_{1\text{u}}$ & 1061 &  & 970 & 0 & 0 &  & 1029 & 0 & 0 &  \\ 
& 17 & e$_{1\text{u}}$ & 1061 &  & 970 & 0 & 0 &  & 1029 & 0 & 0 &  \\ 
& 18 & b$_{1\text{u}}$ & 1017 &  & 999 & 0 & 0 &  & 1006 & 0 & 0 &  \\ 
& 19 & b$_{2\text{g}}$ & 1013 &  & 782 & 0 & 0 &  & 791 & 0 & 0 &  \\ 
& 20 & a$_{1\text{g}}$ & 1013 &  & 957 & -21.28 & -1.022 &  & 1035 & -19.69
& -0.946 &  \\ 
& 21 & e$_{2\text{u}}$ & 983 &  & 746 & 0 & 0 &  & 755 & 0 & 0 &  \\ 
& 22 & e$_{2\text{u}}$ & 983 &  & 746 & 0 & 0 &  & 755 & 0 & 0 &  \\ 
& 23 & e$_{1\text{g}}$ & 863 &  & 594 & 0 & 0 &  & 603 & 0 & 0 &  \\ 
& 24 & e$_{1\text{g}}$ & 863 &  & 594 & 0 & 0 &  & 603 & 0 & 0 &  \\ 
& 25 & b$_{2\text{g}}$ & 712 &  & 332 & 0 & 0 &  & 231 & 0 & 0 &  \\ 
& 26 & a$_{2\text{u}}$ & 689 &  & 555 & 0 & 0 &  & 577 & 0 & 0 &  \\ 
& 27 & e$_{2\text{g}}$ & 619 &  & 529 & 0 & 0 &  & 526 & 0 & 0 &  \\ 
& 28 & e$_{2\text{g}}$ & 619 &  & 529 & 0 & 0 &  & 526 & 0 & 0 &  \\ 
& 29 & e$_{2\text{u}}$ & 412 &  & 236 & 0 & 0 &  & 94 & 0 & 0 &  \\ 
& 30 & e$_{2\text{u}}$ & 412 &  & 236 & 0 & 0 &  & 94 & 0 & 0 &  \\ 
\hline\hline
\end{tabular}%
\end{table}

\section{Validation of the Born-Oppenheimer approximation}

To justify the use of Born-Oppenheimer molecular dynamics, we
consider the energy separation between the excited and ground electronic
states as an indicator of crossings between the two surfaces. Figure~\ref%
{fig:EnergyGap} shows that the energy gap between the two electronic states
is rather large during the propagation, implying the validity of the
Born-Oppenheimer approximation. We note, however, that nonadiabatic
couplings to the higher excited states could cause the breakdown of the
Born-Oppenheimer approximation, especially in the case of the phenyl radical,
where even the on-the-fly approach cannot reproduce the experimental
spectrum perfectly. Finally, let us mention that the energy gap is only one of several criteria for determining the importance of nonadiabatic couplings. A more rigorous criterion monitors the population transfer and the most rigorous criterion monitors directly the \emph{adiabaticity}, i.e., the fidelity between the Born-Oppenheimer and nonadiabatic wavefunctions.\cite{Zimmermann_Vanicek:2010,Zimmermann_Vanicek:2012,Zimmermann_Vanicek:2012a,Bircher_Rothlisberger:2017}

\begin{figure}[H]
\centering
\includegraphics[scale=0.8]{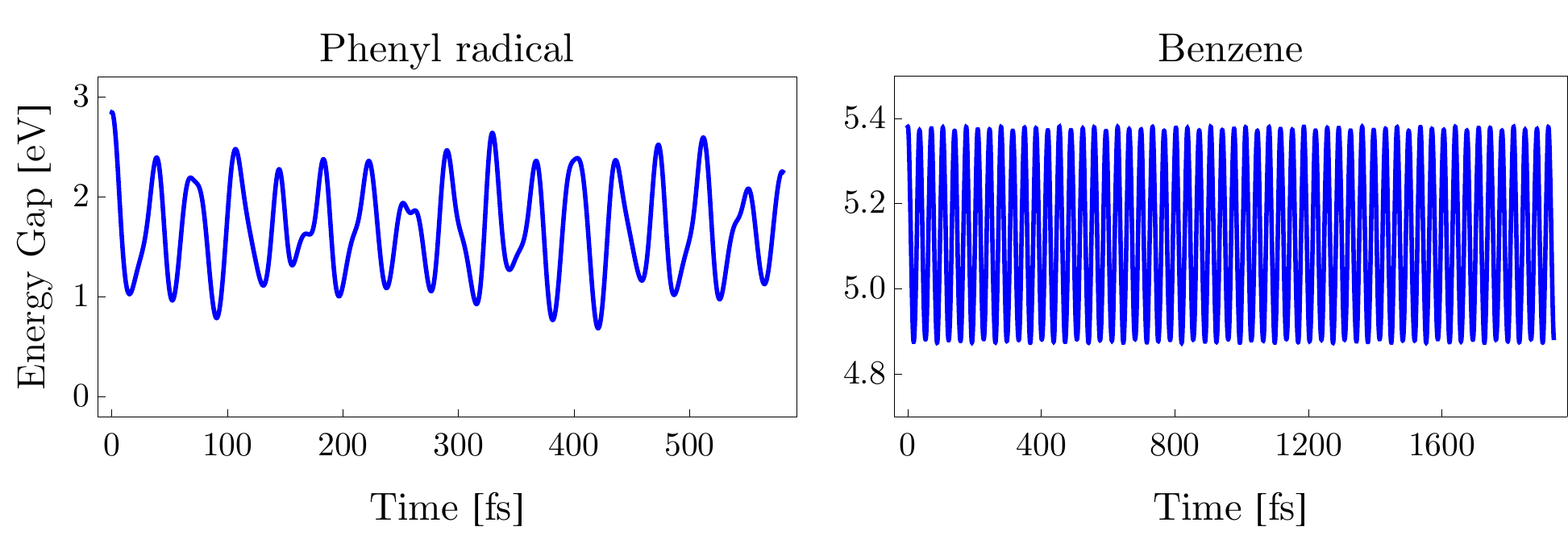}
\caption{The energy gap between the excited and the ground electronic state
for the phenyl radical (left) and benzene (right) evaluated along the \textit{ab
initio} trajectory.}
\label{fig:EnergyGap}
\end{figure}

\section{Validation of the electronic structure method}

\begin{figure}[H]
\centering
\includegraphics[scale=0.8]{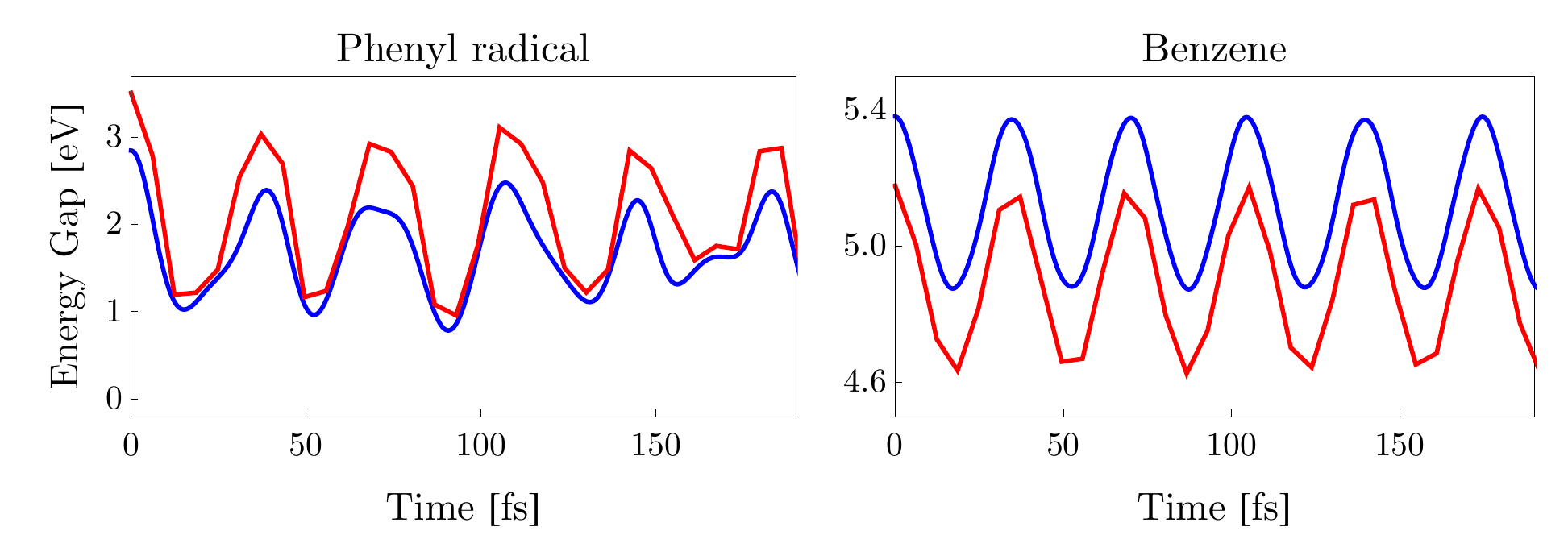}
\caption{As in Fig.~\protect\ref{fig:EnergyGap}, but shown on a shorter time
scale and compared to the energies computed using EOM-CCSD method (red).}
\label{fig:CCSD}
\end{figure}

To verify that the chosen electronic structure methods are reliable
for benzene and the phenyl radical, we show the energies computed using the
equation-of-motion coupled cluster (EOM-CCSD) method along the TDDFT
trajectory (Fig.~\ref{fig:CCSD}). Same basis sets were used as in the TDDFT
calculations: SNSD for the phenyl radical and 6-31+G(d,p) for benzene. The
single point energy calculations were performed every 32 steps of the
trajectory for the initial 200 fs. Although there is a shift in energy
between the two methods, both give similar curvature of the potential energy
surface. Our findings are in accord with the literature; indeed, B3LYP
functional and the two basis sets have already been validated for spectra
calculations of the phenyl radical and benzene by comparison with high-level
electronic structure methods.\cite{Baiardi_Barone:2013,Li_Lin:2010}

\section{Conservation of the phase space volume and symplectic structure}

Classical dynamics preserves phase space volume. This statement,
known as the Liouville theorem, is only one of $D$ corollaries of
the conservation of the symplectic structure, expressed by the equation %
\begin{equation}
M_{t}^{\top }\cdot J\cdot M_{t}=J\,,  \label{eq:symplectic_condition}
\end{equation}%
where $M_{t}$ is the $2D\times 2D$ stability
matrix evaluated at time $t$, 
\begin{equation}
J=%
\begin{pmatrix}
0 & I \\ 
-I & 0%
\end{pmatrix}%
\,
\end{equation}%
is the standard symplectic matrix, $I$ denotes the $D$-dimensional identity matrix, and $D$ the number of
degrees of freedom. The conservation of phase space volume, expressed by the
condition%
\begin{equation}
\det M_{t}=1,
\end{equation}%
follows from Eq.~(\ref{eq:symplectic_condition}). The fact that $%
\det M_{t}=\pm 1$ follows immediately by taking the determinant of
both sides. Of the two options only $\det M_t = +1$ is possible since $\det M_0 = \det I_{2D} = +1$, since the determinant of a matrix is a continuous function, and since the stability matrix is a continuous function of time.

While our semiclassical propagation uses symplectic algorithms, the
numerical implementation is performed at a finite precision; as a result,
the conservation of symplectic structure and phase space volume are not
guaranteed. However, monitoring their conservation provides a useful check
of the semiclassical dynamics. Since the propagation of the thawed Gaussian
wavepacket requires propagating the stability matrix, verifying the
conservation laws is straightforward.

\begin{figure}[H]
\centering
\includegraphics[scale=0.8]{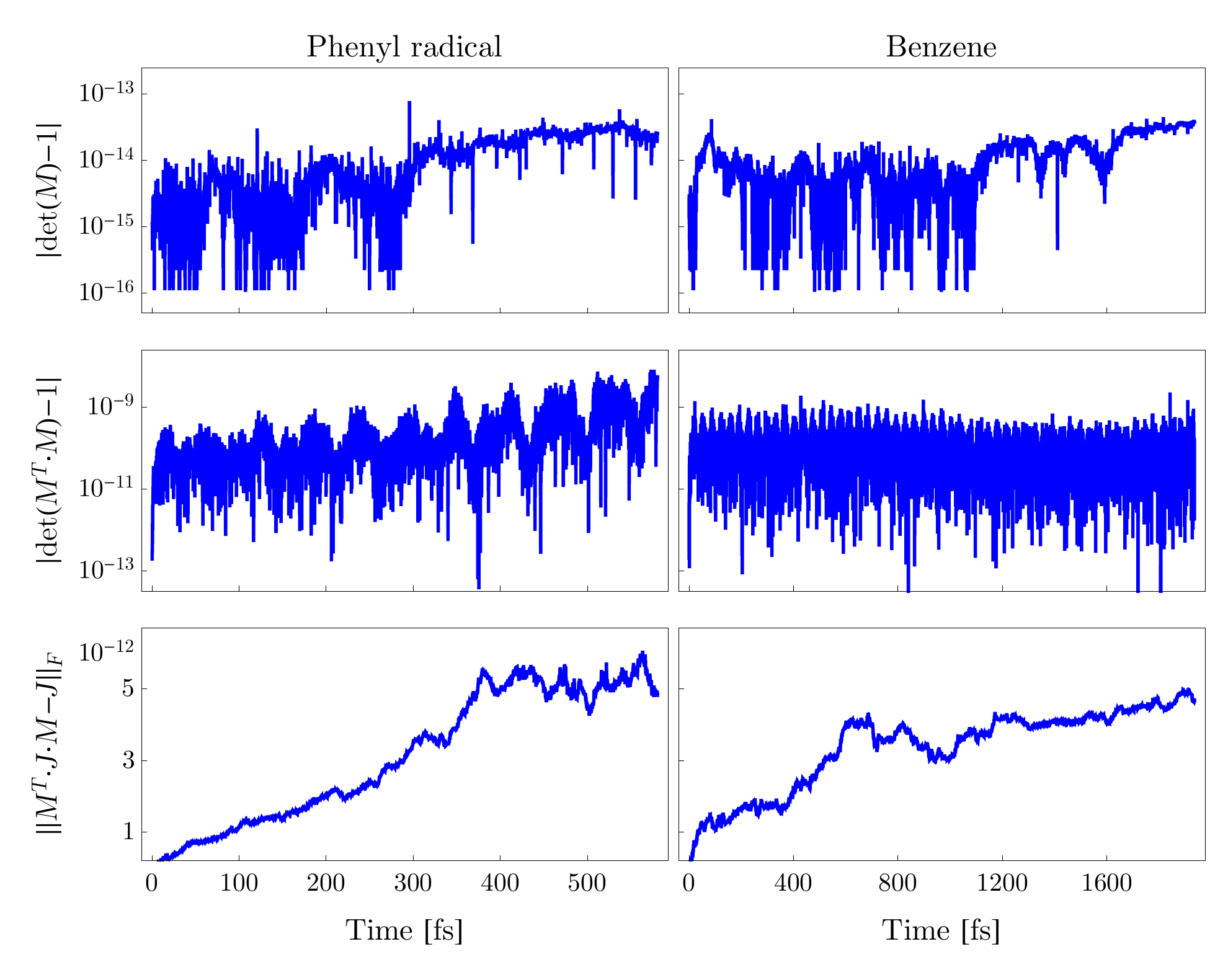}
\caption{The errors in the determinant of the stability matrix (top panels),
the determinant of $M^\top \cdot M$ (middle panels), and the symplectic
condition (bottom panels) for the phenyl radical (left) and benzene (right).}
\label{fig:StabilityMatrix}
\end{figure}

To check the symplectic condition (\ref{eq:symplectic_condition}),
we monitor the Frobenius norm $||\cdot ||_{F}$ of the difference $%
M_{t}^{\top }\cdot J\cdot M_{t}-J$; recall that the Frobenius norm
of a general matrix $A$ is defined as %
\begin{equation}
||A||_{F}:=\sqrt{\text{Tr}(A^{\top }\cdot A)}\,.
\end{equation}%
To check the conservation of phase space volume, we monitor $\det
M_{t}$ and $\det (M_{t}^{\top }\cdot M_{t})$ both of
which should be equal to 1.\cite{Zhuang_Ceotto:2013} The time dependence of
all three quantities is shown in Fig.~\ref{fig:StabilityMatrix}, which
confirms that our semiclassical dynamics conserves both phase space volume
and symplectic structure within machine precision during the time that
determines the spectra of benzene and the phenyl radical. Note, however, that even when
the symplectic structure and phase space volume are not satisfied so
accurately, it is possible to extend the semiclassical propagation time by
regularizing the stability matrix.\cite{DiLiberto_Ceotto:2016}

\section{Horizontal shifts introduced into the calculated phenyl radical and
benzene spectra}

\begin{table}[H]
\caption{Overall frequency shifts (in cm$^{-1}$) introduced into the
calculated spectra of the phenyl radical and benzene. Same shifts were used for
Franck-Condon, Franck-Condon Herzberg-Teller, and ``Franck-Condon'' spectra
(see main text for definition).}
\label{tab:shifts}\centering
\begin{tabular}{ccc}
\hline\hline
& Phenyl radical & Benzene \\ \hline
OTF-AI & -437 & 3010 \\ 
AH & -297 & 3020 \\ 
VH & -267 & 3300 \\ \hline\hline
\end{tabular}%
\end{table}

\section{Models describing the failure of the vertical harmonic method for
computing absorption spectra}

\subsection{Phenyl radical}

\begin{figure}[H]
\centering
\includegraphics[scale=0.8]{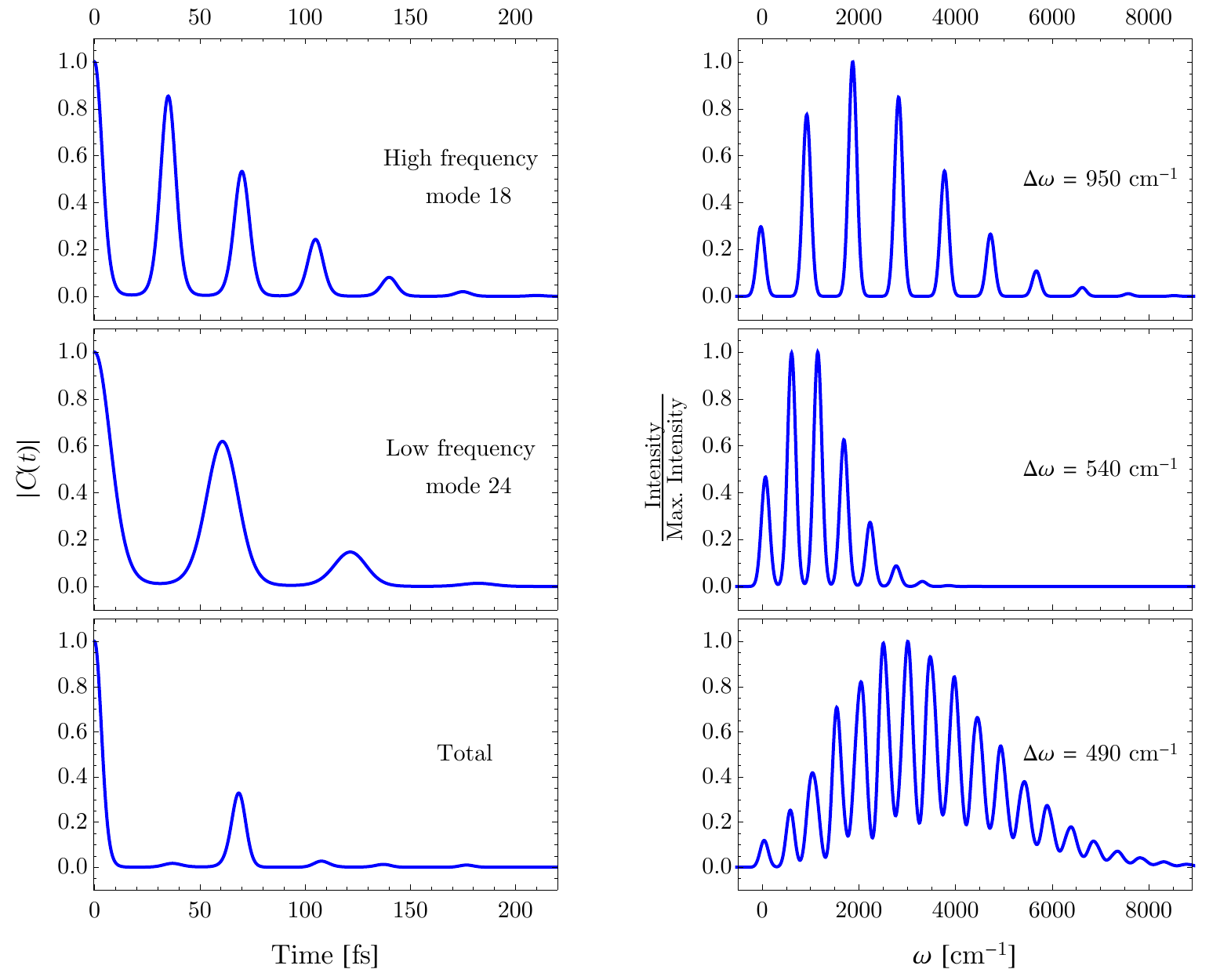}
\caption{Autocorrelation functions (left panels) for a model system
comprising two modes (mode 18 and mode 24) of the phenyl radical, and the
corresponding spectra (right panels). The frequencies used correspond to the
vertical harmonic model of the excited-state potential. When the modes are
not coupled, the product of their autocorrelation functions represents the
total autocorrelation function of the system, i.e., the autocorrelation
function that would be obtained if a two-dimensional simulation were
performed. It can be easily seen that the first small recurrence in the
total autocorrelation function appears at a time that is different from any
of the recurrences in the two one-dimensional cases. This is reflected in
the corresponding spectra, each of which is showing a single progression. 
\emph{While the spacings in the first two progressions are equal to the
corresponding excited-state frequencies of the two modes, the overall
spectrum contains only a single progression with a spacing of 490 cm$^{-1}$,
resembling the situation observed in the vertical harmonic spectrum of
the phenyl radical.} This value corresponds to neither of the two frequencies of
the system and is called the \emph{missing mode effect} frequency.}
\label{fig:MIME}
\end{figure}

\subsection{Benzene}

\begin{figure}[H]
\centering
\includegraphics[scale=1]{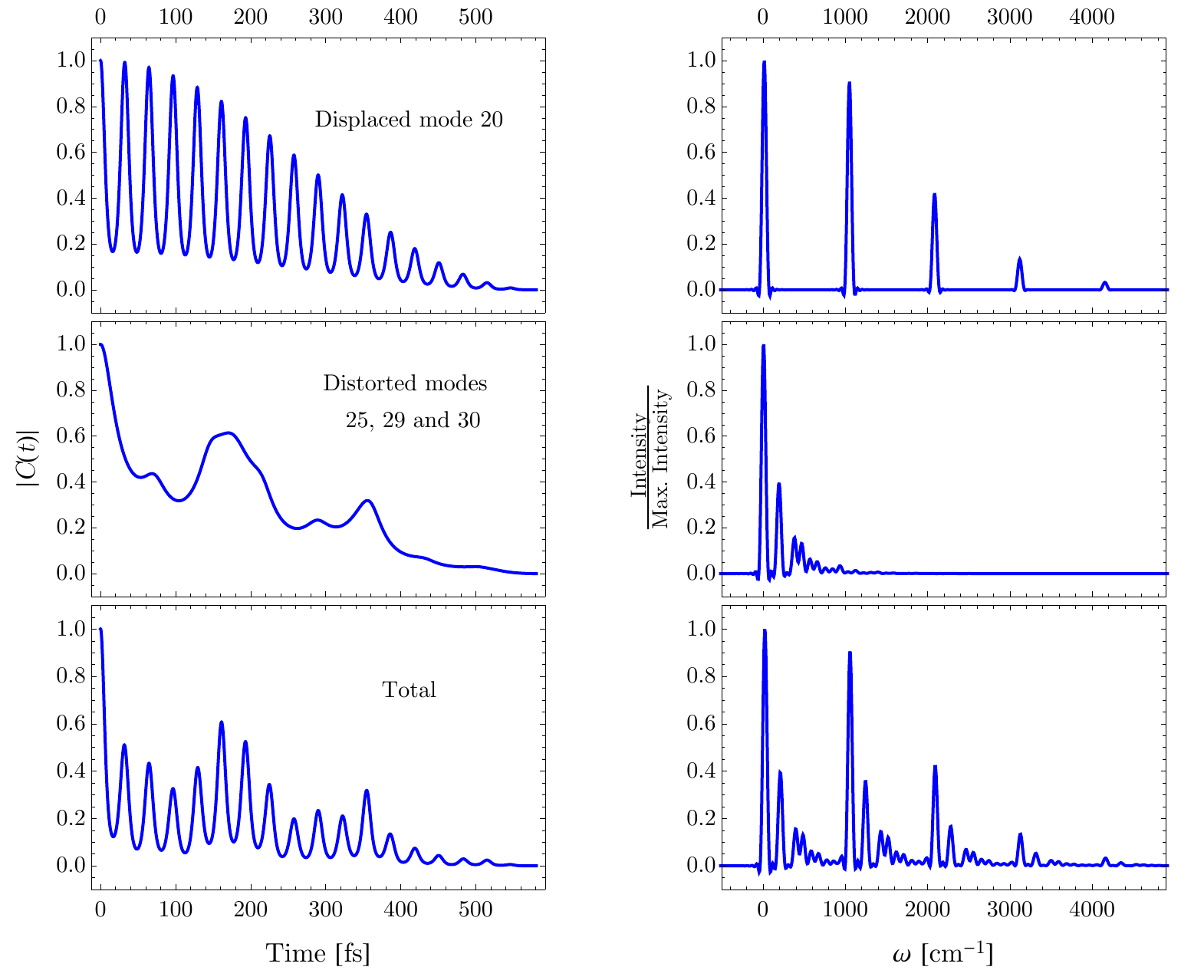}
\caption{A four-dimensional model with three undisplaced highly distorted
low-frequency modes, which correspond to benzene excited-state normal modes
25, 29 and 30, and one displaced mode, corresponding to the mode 20, is used
to explain the observations in the vertical harmonic spectrum of benzene.
The autocorrelation functions for the single displaced mode and the three
distorted modes are given on the left, with the resulting spectra on the
right. The total autocorrelation function (bottom, left) is the product of
the two autocorrelation functions, and the corresponding spectrum is
therefore a convolution of the two spectra in the frequency domain (bottom,
right). \emph{The resulting spectrum for this four-dimensional model is
indeed similar to the observed VH spectrum and proves that overestimated
distortions, i.e., incorrect low frequencies of the modes 25, 29 and 30,
result in having more peaks than observed in the experimental spectrum.}}
\label{fig:1Dx3D_Final}
\end{figure}

\section{Anharmonicity effects in benzene}

The VH model results in a qualitatively wrong absorption spectrum of
benzene. In contrast, the static adiabatic harmonic picture is useful to
describe the spectrum qualitatively, but does not provide correct
intensities. A more dynamical picture is obtained by representing the
vibrational motion in phase space. Significant dynamics is present only in
the normal mode 20, whereas the other modes exhibit almost no classical
dynamics. It is important to recall here \emph{that the positions and momenta in the OTF approach result
only from an on-the-fly \textit{ab initio} classical dynamics on the excited
state surface and do not depend on the accuracy of the excited-state Hessian
calculations, while the dynamics in the global harmonic methods depends
solely on the Hessian and the displacement of the excited-state potential.}
This results in the differences observed in the phase space (see Fig.~\ref%
{fig:AH_modified_Benzene}, top left).

\begin{figure}[pth]
\centering \includegraphics[scale=1]{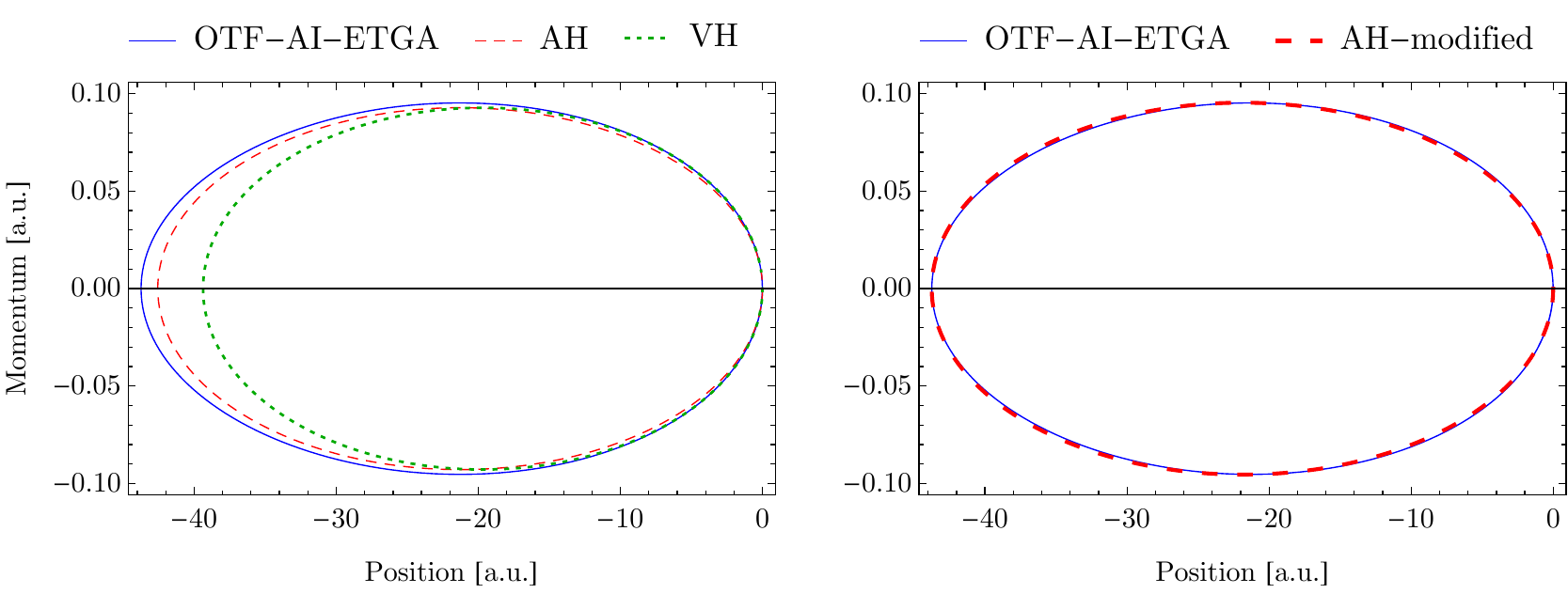} \newline
\vspace{0.5 cm} \centering %
\includegraphics[scale=1]{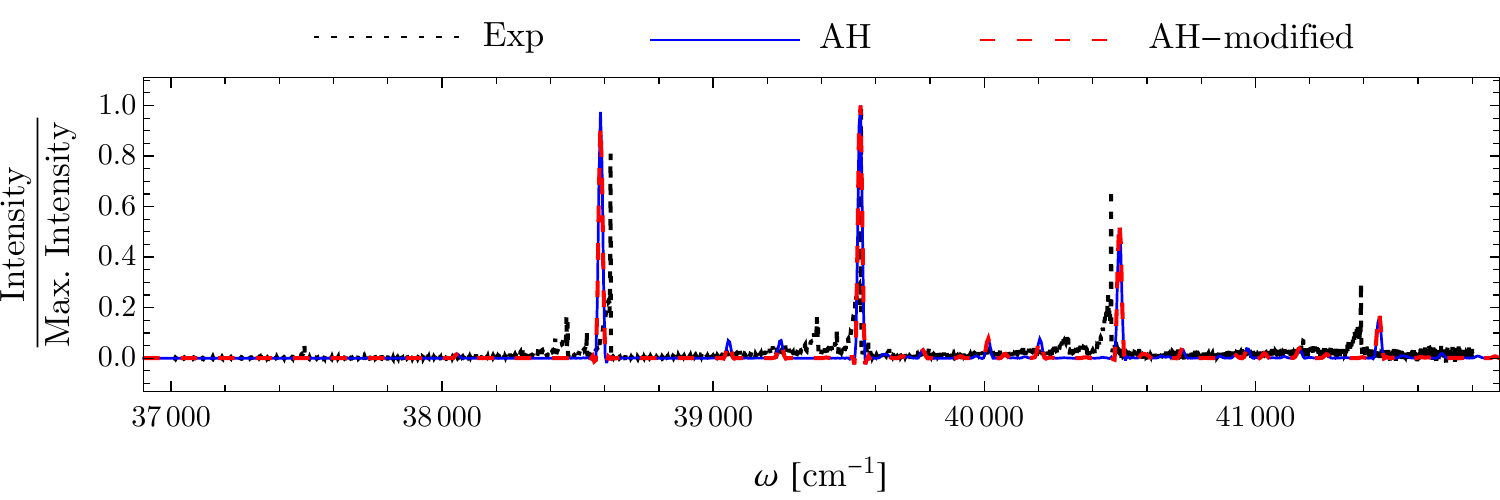}
\caption{Phase-space representation of the dynamics in the normal mode 20 of
benzene (top panels) and the spectrum (bottom panel) obtained using a
modified adiabatic harmonic potential (AH-modified). Position and momentum
are given in the mass-scaled normal mode coordinates in atomic units (1~a.u.~%
$\simeq 0.0123943 \,\protect\sqrt{\text{amu}}\,\text{\AA }$ for position and
1~a.u.~$\simeq 0.512396 \,\protect\sqrt{\text{amu}}\,\text{\AA } / \text{{fs}%
}$ for momentum). The top left panel shows trajectories corresponding to
three different methods---the on-the-fly, adiabatic harmonic, and vertical
harmonic propagations, while the top right panel compares the on-the-fly
classical dynamics with the dynamics in the modified adiabatic harmonic
potential, i.e., the potential obtained by setting the displacement of the
adiabatic harmonic potential in the normal mode coordinate 20 to the
displacement observed in the OTF-AI dynamics. The AH-modified spectrum is
compared to the original AH and experimental spectra in the bottom panel.
Despite the similarity between the on-the-fly dynamics and the dynamics in
the AH-modified potential (top right panel), the modified adiabatic harmonic
spectrum shows no significant improvement over the original AH spectrum.}
\label{fig:AH_modified_Benzene}
\end{figure}

The differences in the dynamics of the phase space center of the wavepacket
can have strong effect on the resulting spectra, both in terms of the
positions and the intensities of the peaks. In benzene, the phase-space
dynamics in Fig.~\ref{fig:AH_modified_Benzene} reveals that the on-the-fly
and adiabatic harmonic approaches are induced by different excited-state
displacements of the normal mode 20. If the wrong displacement of the AH
potential were the cause of poor spectral intensities, changing it to the
value of the displacement of the OTF calculation would improve the spectrum.
To test this hypothesis, a ``modified'' adiabatic harmonic potential is
constructed by setting the displacement in the normal mode 20 to the
displacement observed in the on-the-fly classical dynamics. However, this
modified adiabatic potential yields a spectrum similar to the original one
(see Fig.~\ref{fig:AH_modified_Benzene}, bottom), even though the classical
motion in the modified AH and OTF potentials energy surfaces are almost the
same (Fig.~\ref{fig:AH_modified_Benzene}, top right). \emph{Thus, the
difference in the classical dynamics is not the reason for the observed
differences in the calculated spectra,} leaving the possibility that the
differences in the spectra can only be explained by considering the width
matrix and the phase of the wavepacket.

\begin{figure}[pth]
\centering \includegraphics[scale=1]{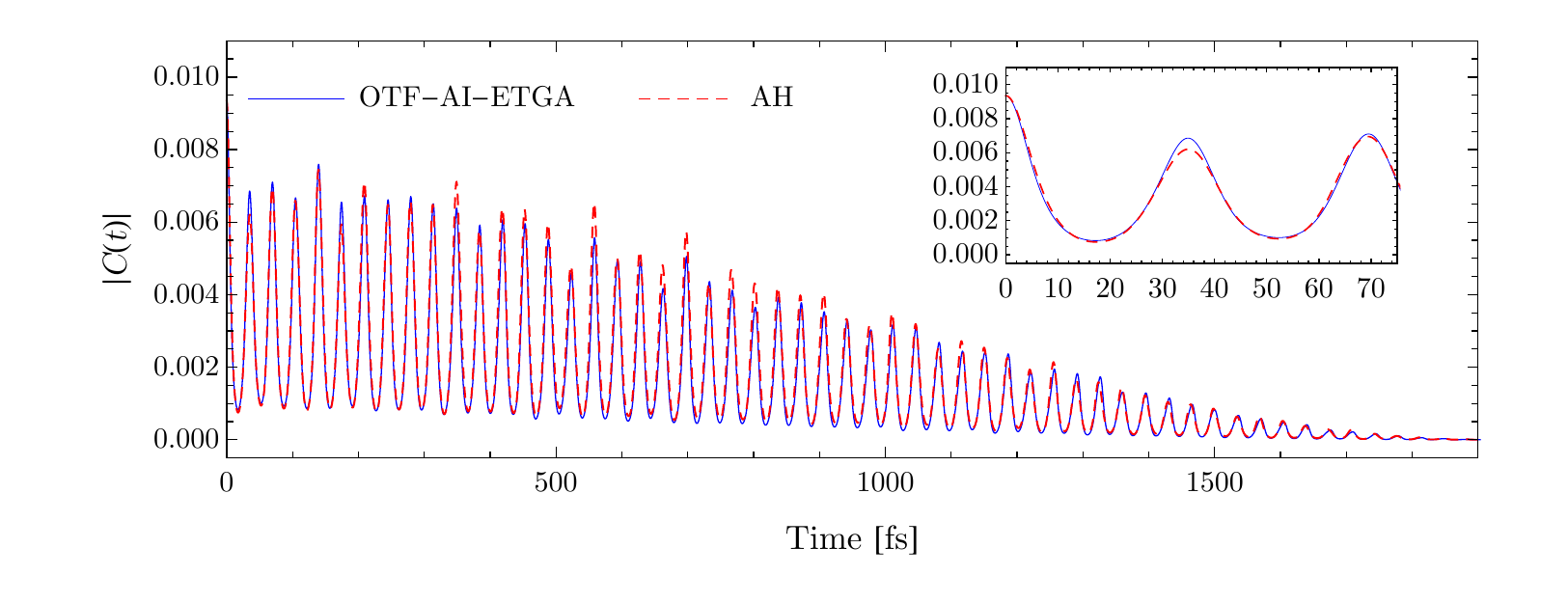} \newline
\centering  \includegraphics[scale=1]{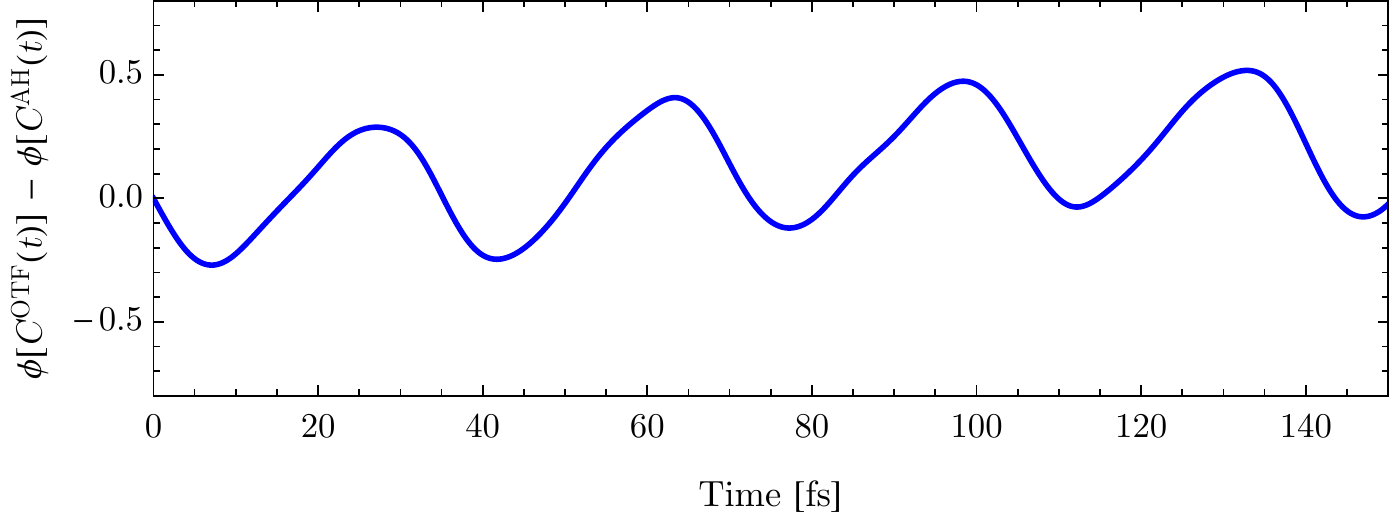}
\caption{Autocorrelation functions calculated using the on-the-fly and
adiabatic harmonic approaches are compared: the absolute values are shown in
the top panel, while the difference of the phases on a shorter timescale is
given in the bottom panel. Only slight differences in the shapes of the
recurrences are observed in the magnitudes of the autocorrelation functions,
while the difference of the phases is significant.}
\label{fig:Benzene_corr}
\end{figure}

Although the classical dynamics is not affected by the anharmonicity of the
excited-state potential, resulting in equally spaced peaks in the spectra,
the intensities of the peaks are influenced by the changes in the
autocorrelation function (Fig.~\ref{fig:Benzene_corr}). Such pattern in the
relative intensities cannot be described by a simple global harmonic
approach, but requires a method that can treat anharmonicity. The agreement
with the experimental spectrum proves that running an on-the-fly trajectory
with Hessians evaluated along the trajectory gives a reliable wavepacket
propagation and accounts for fine details in the autocorrelation function.

\begin{figure}[pth]
\centering
\includegraphics[scale=1]{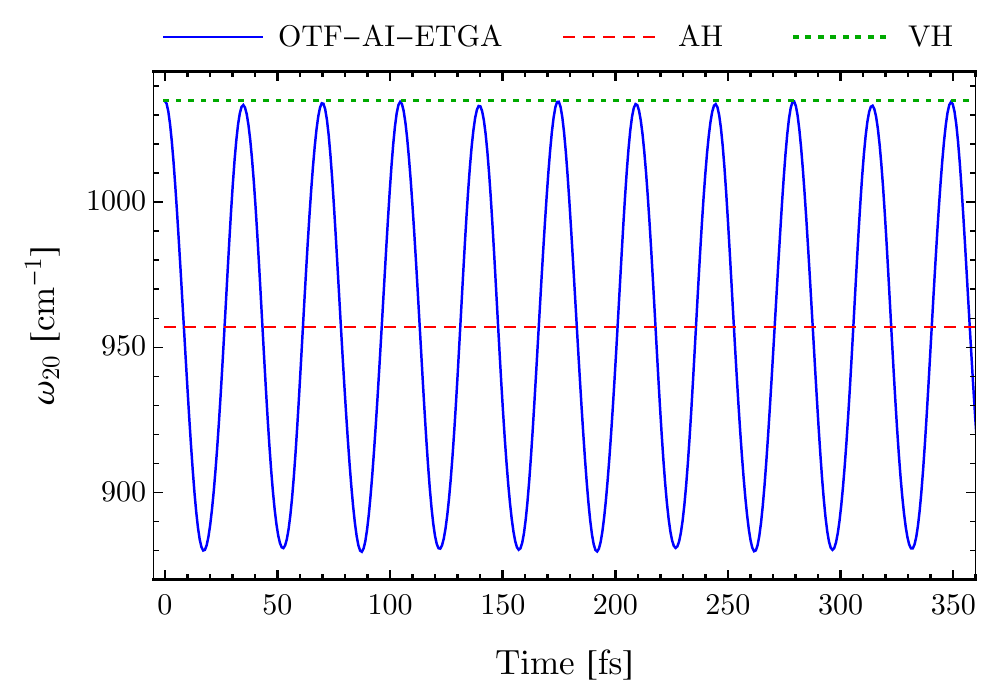}
\caption{Frequency of the normal mode 20 of benzene in the excited
electronic state as a function of time. Comparison of the on-the-fly,
adiabatic harmonic, and vertical harmonic approaches.}
\label{fig:freqBenzene}
\end{figure}

To further investigate the differences between the adiabatic harmonic
approximation and the OTF-AI-ETGA, we explore the dynamics in the most
active totally symmetric normal mode 20. Vertical and adiabatic harmonic
Hessians yield different frequencies of this mode and both approaches (see
Table~\ref{tab:Benzene_freq}) propagate the wavepacket using only one value
for the frequency of the mode (by definition, since they both use global
harmonic potentials). However, the true potential is not perfectly harmonic. 
\emph{The stiffness of the mode 20 changes along the trajectory, with the
vertical Hessian representing the maximum and the adiabatic Hessian giving a
value roughly equal to the mean frequency (see Fig.~\ref{fig:freqBenzene}),
explaining the better agreement of the AH with the OTF approach.}

\section{``Franck-Condon'' (``FC'') spectrum and absolute absorption cross
sections of benzene}

\begin{figure}[H]
\centering
\includegraphics[scale=1]{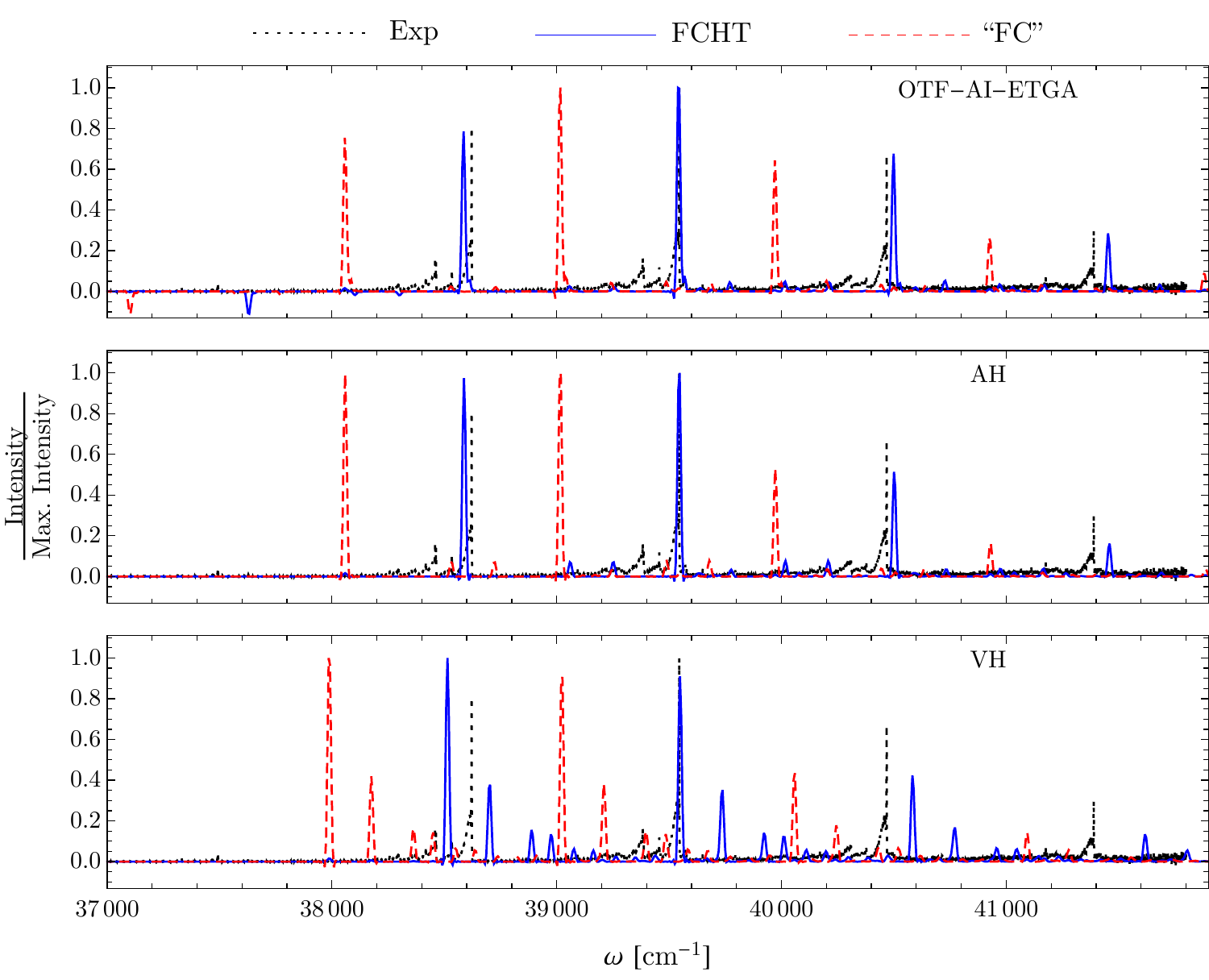}
\caption{``FC'' and FCHT spectra of benzene evaluated using the adiabatic
harmonic (AH), vertical harmonic (VH), and on-the-fly \textit{ab initio}
extended TGA (OTF-AI-ETGA) approaches are compared with the experimental
spectrum. All spectra were horizontally shifted and scaled according to the
highest peak of the FCHT calculated spectra (see Table~\protect\ref%
{tab:shifts}). The same horizontal shift for both ``FC'' and FCHT spectra
was used as to show that the constant horizontal shift between them
corresponds to the excited-state frequency of the inducing modes 27 and 28: $%
\protect\omega = 529 \,\text{cm}^{-1}$ for the AH, $\protect\omega = 526 \,%
\text{cm}^{-1}$ for the VH model.}
\label{fig:Benzene_shiftedFC}
\end{figure}

\begin{figure}[H]
\centering
\includegraphics[scale=0.9]{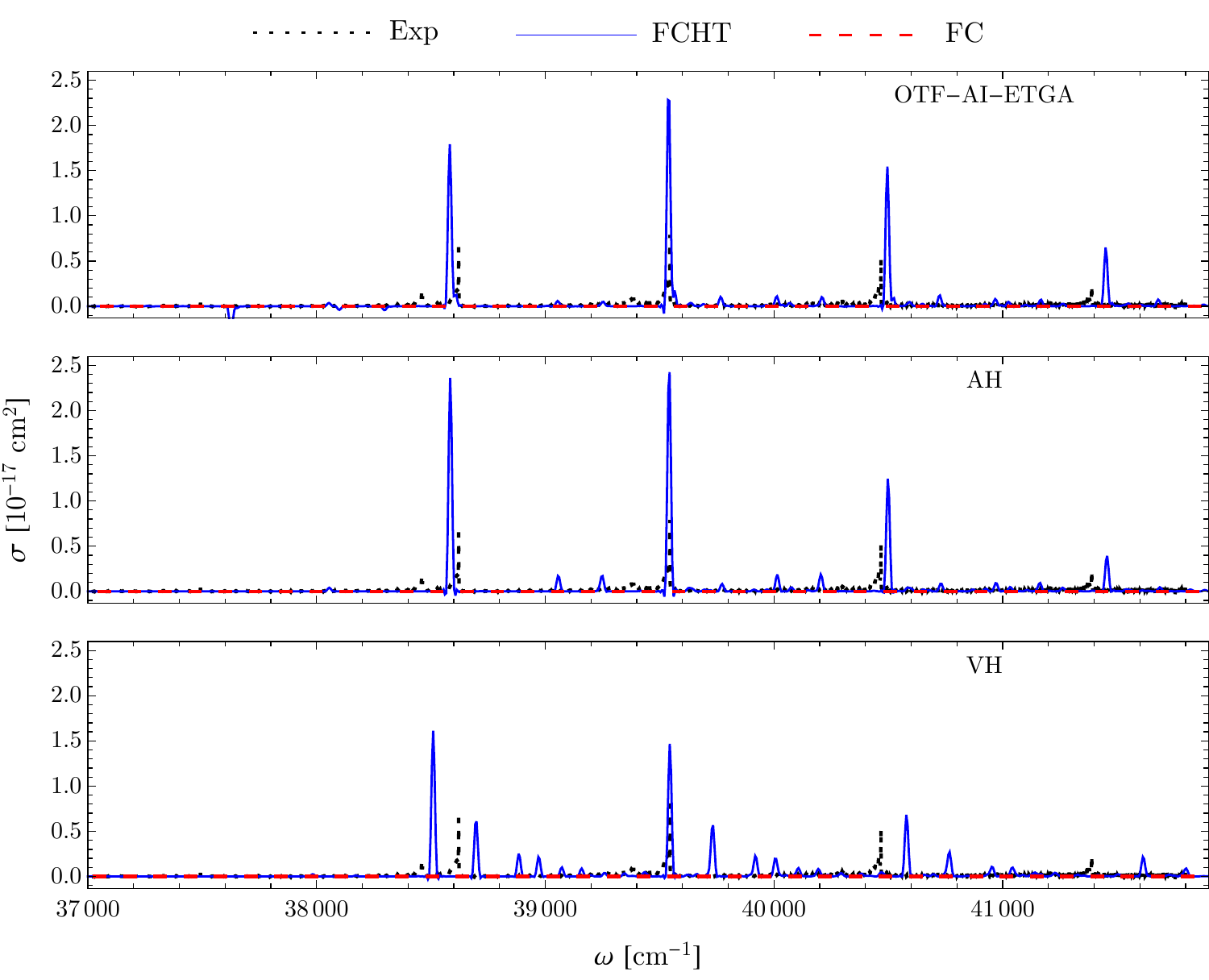}
\caption{Absolute absorption cross sections, evaluated with the on-the-fly
and global harmonic approaches, are compared to the experimental spectrum.
The calculated spectra were horizontally shifted as in Fig.~\protect\ref%
{fig:Benzene_shiftedFC}, but the magnitudes were not scaled. While the FC
spectra are zero, because the transition dipole moment is zero, the spectra
computed within the Herzberg-Teller approximation (FCHT) are in fair
agreement with the experiment, although significantly overestimating the
experimental values by a roughly constant systematic factor of $\approx 3$.}
\label{fig:Benzene_absolute}
\end{figure}

\bibliography{Herzberg-Teller_TGA}